\documentstyle[12pt,bezier]{article}

\makeatletter

\def\section{\@startsection {section}{1}{\z@}{+3.0ex plus +1ex minus
  +.2ex}{2.3ex plus .2ex}{\normalsize\bf}}
\def\subsection{\@startsection{subsection}{2}{\z@}{+2.5ex plus +1ex
minus +.2ex}{1.5ex plus .2ex}{\normalsize\bf}}
\def\subsubsection{\@startsection{subsubsection}{3}{\z@}{+3.25ex plus
 +1ex minus +.2ex}{1.5ex plus .2ex}{\normalsize\bf}}

\def\appendix{\setcounter{section}{0} \setcounter{subsection}{0}
              \setcounter{equation}{0}
              \def\thesection{\Alph{section}}
              \def\theequation{\thesection\arabic{equation}}}
\makeatletter

\newcount\@tempcntc
\def\@citex[#1]#2{\if@filesw\immediate\write\@auxout{\string\citation{#2}}\fi
  \@tempcnta\z@\@tempcntb\m@ne\def\@citea{}\@cite{\@for\@citeb:=#2\do
    {\@ifundefined
       {b@\@citeb}{\@citeo\@tempcntb\m@ne\@citea
        \def\@citea{,\penalty\@m\ }{\bf ?}\@warning
       {Citation `\@citeb' on page \thepage \space undefined}}%
    {\setbox\z@\hbox{\global\@tempcntc0\csname
b@\@citeb\endcsname\relax}%
     \ifnum\@tempcntc=\z@ \@citeo\@tempcntb\m@ne
       \@citea\def\@citea{,\penalty\@m}
       \hbox{\csname b@\@citeb\endcsname}%
     \else
      \advance\@tempcntb\@ne
      \ifnum\@tempcntb=\@tempcntc
      \else\advance\@tempcntb\m@ne\@citeo
      \@tempcnta\@tempcntc\@tempcntb\@tempcntc\fi\fi}}\@citeo}{#1}}

\def\@citeo{\ifnum\@tempcnta>\@tempcntb\else\@citea
  \def\@citea{,\penalty\@m}%
  \ifnum\@tempcnta=\@tempcntb\the\@tempcnta\else
   {\advance\@tempcnta\@ne\ifnum\@tempcnta=\@tempcntb \else
\def\@citea{--}\fi
    \advance\@tempcnta\m@ne\the\@tempcnta\@citea\the\@tempcntb}\fi\fi}

\expandafter\ifx\csname mathrm\endcsname\relax\def\mathrm#1{{\rm #1}}\fi

\def\no{\nonumber\\}

\def\beq{\begin{equation}}
\def\eeq{\end{equation}}
\def\beqar{\begin{eqnarray}}
\def\eeqar{\end{eqnarray}}
\def\barr#1{\begin{array}{#1}}
\def\earr{\end{array}}
\def\disp{\displaystyle}
\def\arraystretch{1.2}
\def\bma{\begin{displaymath}}
\def\ema{\end{displaymath}}

\def\be{\beta}
\def\Ga{\Gamma}
\def\ga{\gamma}
\def\de{\delta}

\def\eps{\epsilon}
\def\veps{\varepsilon}
\def\la{\lambda}
\def\om{\omega}
\def\si{\sigma}
\def\Si{\Sigma}

\def\refeq#1{\mbox{(\ref{#1})}}
\def\refeqs#1{\mbox{(\ref{#1})}}
\def\refeqf#1{\mbox{(\ref{#1})}}
\def\refse#1{\mbox{section~\ref{#1}}}
\def\citere#1{\mbox{Ref.~\cite{#1}}}
\def\citeres#1{\mbox{Refs.~\cite{#1}}}

\renewcommand{\O}{{\cal{O}}}
\renewcommand{\L}{{\cal{L}}}

\def\mathswitchr#1{\relax\ifmmode{\mathrm{#1}}\else$\mathrm{#1}$\fi}
\newcommand{\Pf}{\mathswitch  f}
\newcommand{\PW}{\mathswitchr W}
\newcommand{\PZ}{\mathswitchr Z}
\newcommand{\PA}{\mathswitchr A}
\newcommand{\PH}{\mathswitchr H}
\newcommand{\Pb}{\mathswitchr b}

\newcommand{\Ff}{\mathswitch f}
\newcommand{\Ffbar}{\mathswitch{\bar f}}

\newcommand{\FZ}{\mathswitch Z}

\newcommand{\FH}{\mathswitch H}

\def\mathswitch#1{\relax\ifmmode#1\else$#1$\fi}
\newcommand{\Mf}{\mathswitch {m_\Pf}}
\newcommand{\MW}{\mathswitch {M_\PW}}
\newcommand{\MA}{\mathswitch {M_\PA}}
\newcommand{\MZ}{\mathswitch {M_\PZ}}
\newcommand{\MH}{\mathswitch {M_\PH}}

\newcommand{\rw}{{\mathrm{W}}}
\newcommand{\sw}{\mathswitch {s_{\rw}}}
\newcommand{\cw}{\mathswitch {c_{\rw}}}

\newcommand{\Qf}{\mathswitch {Q_\Pf}}
\newcommand{\vf}{\mathswitch {v_\Pf}}
\newcommand{\af}{\mathswitch {a_\Pf}}

\def\asymp#1{\mathrel{\raisebox{-.4em}{$\widetilde{\scriptstyle #1}$}}}
\def\ie{i.e.\ }
\newcommand{\se}{self-energy}
\newcommand{\ses}{self-energies}

\newcommand{\WI}{Ward identities}

\newsavebox{\Vr}
\newsavebox{\Sr}
\newsavebox{\Fr}
\newsavebox{\Vtr}
\newsavebox{\Vbr}
\newsavebox{\Str}
\newsavebox{\Sbr}
\newsavebox{\Ftr}
\newsavebox{\Fbr}
\newsavebox{\Vtbr}
\newsavebox{\Ftbr}
\newsavebox{\Stbr}
\newsavebox{\Gr}
\newsavebox{\Gtbr}
\newsavebox{\Vp}
\newsavebox{\wigr}
\newsavebox{\wigur}
\newsavebox{\wigdr}

\def\Dhat{\hat D}
\def\Vhat{\hat V}
\def\What{\hat W}
\def\Bhat{\hat B}
\def\Zhat{\hat Z}
\def\Ahat{\hat A}
\def\Shat{\hat S}
\def\Hhat{\hat H}
\def\Phihat{\hat \Phi}
\def\phihat{\hat \phi}
\def\chihat{\hat \chi}
\def\thetahat{\hat \theta}
\def\xiQ{\xi_Q}
\def\rY{{\mathrm{Y}}}
\def\rT{{\mathrm{T}}}
\def\rS{{\mathrm{S}}}
\newcommand{\rL}{\mathswitchr L}
\newcommand{\rR}{\mathswitchr R}
\def\ps{\! \! \not \! p \,}
\def\ks{\! \! \not \! k \,}
\newcommand{\ren}{\mathrm{ren}}
\newcommand{\dive}{\mathrm{div}}
\renewcommand{\Re}{\mathop{\mathrm{Re}}}

\newtoks\@stequation

\def\subequations{\refstepcounter{equation}%
  \edef\@savedequation{\the\c@equation}%
  \@stequation=\expandafter{\theequation}%   %only want \theequation
  \edef\@savedtheequation{\the\@stequation}% %expanded once
  \edef\oldtheequation{\theequation}%
  \setcounter{equation}{0}%
  \def\theequation{\oldtheequation\alph{equation}}}%

\def\endsubequations{%
  \setcounter{equation}{\@savedequation}%
  \@stequation=\expandafter{\@savedtheequation}%
  \edef\theequation{\the\@stequation}\global\@ignoretrue}

\begin{document}

\thispagestyle{empty}
\def\thefootnote{\fnsymbol{footnote}}
\setcounter{footnote}{1}
\null
\renewcommand{\baselinestretch}{1}
\Huge
\normalsize
\mbox{} \hfill BI-TP. 94/50\\
\mbox{} \hfill UWITP 94/03\\
\mbox{} \hfill hep-ph/9410338
\vskip 1cm
\vfill
\begin{center}
{\Large \bf
\boldmath{Application of the Background-Field Method
to the electroweak Standard Model}
\par} \vskip 2.5em
{\large
{\sc Ansgar Denner, Georg Weiglein%
\footnote{E-mail: weiglein@vax.rz.uni-wuerzburg.de}
} \\[1ex]
{\normalsize \it Institut f\"ur Theoretische Physik, Universit\"at W\"urzburg\\
Am Hubland, D-97074 W\"urzburg, Germany}
\\[2ex]
{\sc Stefan Dittmaier%
\footnote{Supported by the Bundesministerium f\"ur Forschung und
Technologie, Bonn, Germany.} \\[1ex]
{\normalsize \it Theoretische Physik, Universit\"at Bielefeld\\
Universit\"atsstra{\ss}e, D-33501 Bielefeld, Germany}
}
\par} \vskip 1em
\end{center} \par
\vskip 2cm %4cm
\vfil
{\bf Abstract:} \par
Application of the background-field method yields a
gauge-invariant effective action for the electroweak Standard
Model, %From the effective action
from which simple QED-like Ward
identities are derived. As a consequence of these Ward identities,
the background-field Green functions are shown to possess very desirable
theoretical properties. The renormalization of the Standard
Model in the background-field formalism is studied. A consistent
on-shell renormalization procedure retaining the full gauge symmetry is
presented.
The structure of the counterterms is shown to greatly simplify
compared to the conventional formalism.
A complete list of Feynman rules
for the Standard Model in the background-field method
is given for
arbitrary values of a quantum gauge parameter including  all counterterms
necessary for one-loop calculations.

\par
\vskip 1cm %2cm
\noindent BI-TP. 94/50 \par
\noindent UWITP 94/03 \par
\vskip .15mm
\noindent October 1994 \par
\null
\setcounter{page}{0}
\clearpage
\def\thefootnote{\arabic{footnote}}
\setcounter{footnote}{0}

\section{INTRODUCTION}

The current theoretical understanding of elementary particle
physics is based on gauge theories, which are constructed following
the principle of gauge invariance. While the classical Lagrangian
is manifestly gauge-invariant, one is forced to fix a gauge in
order to quantize the theory.
In the conventional formulation, %this %gauge-fixing
the gauge symmetry is spoiled in intermediate steps of calculations
and can only be restored at the very end
by projecting on physical degrees of freedom.

To avoid the explicit breaking of gauge symmetry, the
background-field method (BFM) \cite{BFMref,Ab81} was developed.
By decomposing the usual gauge field into a quantum field and a
background field one can impose the gauge fixing necessary for
quantization while keeping the gauge invariance of the
effective action.
The BFM proved to be a valuable tool in gauge theories facilitating
computations both technically and conceptually.
It has found many applications in gravity and supergravity~\cite{grav}
and also in QCD, e.g.~for the calculation of
the $\be$-function~\cite{Ab81,QCDren}.
The equivalence of the S matrix in the BFM to the conventional
one has been proven in~\citere{Ab83}.
In the recent formulation of string motivated rules for more
efficient
computations in gauge theories, the BFM plays an important
role~\cite{string}.
An application to the electroweak one-loop process $\PZ
\rightarrow 3 \ga$ was presented in~\citere{Be93}.
The advantages of calculating S-matrix elements within the BFM
are mainly due to the fact that the gauge fixing of the
background fields is completely independent of the quantum
gauge fixing. The choice of an appropriate background gauge
can simplify practical calculations considerably.
However, for spontaneously broken gauge theories the BFM has hardly been
used. There exists no complete formulation of the BFM for the
electroweak Standard Model (SM). In particular, the renormalization
has not been worked out in detail.

Recently it was shown~\cite{bgf,bgfproc} that application of the BFM
in QCD and the electroweak SM yields Green
functions with very desirable theoretical properties.
They fulfill simple QED-like \WI\ and, in comparison to their
counterparts in the conventional $R_{\xi}$-gauge formalism, often
have an improved asymptotic, UV, and IR behavior.
The issue of obtaining Green functions with suitable
properties has found considerable interest in the literature
during the last years~\cite{Ke89,Ku91,pinch,sirlin}. It is especially
important
for applications dealing with off-shell Green functions.
These become relevant when higher-order contributions are
resummed in order to define running coupling constants
or to take into account finite-width effects in resonance
regions. Furthermore, off-shell formfactors are frequently
discussed, e.g.~for the neutrino or for the top quark.
Off-shell \ses\ are often used to parametrize electroweak
radiative corrections.

Most previous attempts for the construction of Green functions
suitable for these purposes aimed on eliminating their gauge-parameter
dependence within a special class of gauges, usually the
$R_{\xi}$ gauges.
To this end new ``Green functions'' were constructed by
rearranging contributions between \ses, vertex and box diagrams. In
particular, the pinch technique (PT)~\cite{pinch,sirlin}
provides a definite prescription for obtaining gauge-parameter
independent quantities at one-loop order. They were found to
fulfill simple %QED-like
\WI\ and possess other
desirable theoretical
properties. Despite these successes, there are a number of
problems related to the PT approach.
The extension of the PT to higher orders is
rather involved~\cite{pin2l}, and even at one-loop order the PT
is not applicable in a straightforward manner to all possible
Green functions. In addition to these technical difficulties,
the PT has also conceptual problems.
Strictly speaking, the resulting building blocks of the S matrix should
not be called Green functions since their field theoretical meaning
has not been clarified.
The process independence of the new ``Green functions''
constructed within the PT has not been proven.
Moreover, the simple \WI\ and other desirable features have not
been derived within the PT but only verified for specific
one-loop examples.

In~\citeres{bgf,bgfproc} it was shown that on the basis of the BFM
these theoretical problems are resolved. The results obtained
within the PT in QCD and the SM were shown to coincide with the
special case
$\xiQ = 1$ of the BFM results, where $\xiQ$ is a quantum gauge
parameter associated with the gauge fixing of the
quantum fields\footnote{The agreement between the BFM results for $\xi_Q
= 1$ obtained in QCD and the corresponding PT
results was also noted in~\citere{hashi}.}.
The BFM vertex functions are directly
derived from the effective action in all orders of perturbation
theory and are evidently process-independent.
The validity of QED-like Ward identities is a direct
consequence of the gauge invariance of the effective action.
Furthermore, one can show that the \WI\ of the BFM
directly imply other desirable properties of the Green
functions.

{}From the formulation of the BFM it follows that the Ward
identities and the desirable features of Green functions
hold for all values of the quantum gauge parameter $\xiQ$.
This fact is of importance in view of the former
treatments~\cite{Ke89,Ku91,pinch,sirlin} which focus on the
elimination of the gauge-parameter dependence. The analysis in
the BFM shows that not the requirement of gauge-parameter
independence is the criterion leading to Green functions with
suitable properties but the \WI\ following from gauge
invariance. The ambiguity of the vertex functions quantified in
the BFM by their dependence on $\xiQ$ is also inherent in the
former treatments where it corresponds to the ambiguity in
choosing different prescriptions for eliminating the gauge-parameter
dependence.

Owing to the aforementioned properties, the BFM is a well suited
formalism for applications in the electroweak SM concerning both the
discussion of off-shell quantities and a technically and
conceptually simplified evaluation of S-matrix elements.
The purpose of this paper is to provide the tools necessary for
applying the BFM in the SM and to investigate consequences
of the explicit gauge invariance present in the BFM formulation.
In particular, an explicit on-shell renormalization of the SM
in the BFM is
worked out in accordance with the gauge invariance of the effective
action.
The gauge invariance implies relations between the renormalization
constants for parameters and fields and greatly
simplifies the renormalization.

The outline of the paper is as follows.
In section~2 we write out the classical Lagrangian in order
to define our conventions and perform the quantization of the SM in the BFM.
The properties of the resulting
gauge-invariant effective action and the construction of the S matrix
are discussed. In section~3
we derive the \WI\ of the
theory. For several examples the differences to the conventional
formalism are discussed.
In section~4 the renormalization of the SM in the BFM is
worked out.
Section~5 illustrates how desirable properties of
the BFM vertex functions can directly be related to the \WI.
In the appendix, a complete list of Feynman rules for the SM
in the BFM is given for an arbitrary value of the quantum gauge
parameter. All counterterms necessary for one-loop calculations
are included.

\section{THE GAUGE-INVARIANT EFFECTIVE ACTION FOR THE STANDARD MODEL}

\subsection{The classical Lagrangian}

In order to define the relevant quantities, we begin with
the classical Lagrangian ${\cal L}_{\mathrm{C}}$ of the
(minimal) electroweak SM. It consists of the Yang-Mills, the
Higgs and the fermion part
\beq
{\cal L}_{\mathrm{C}} = {\cal L}_{\mathrm{YM}}
+ {\cal L}_{\mathrm{H}} + {\cal L}_{\mathrm{F}}  .
\label{LC}
\eeq
The Yang-Mills part is given as
\beq
{\cal L}_{\mathrm{YM}} = -\frac{1}{4} \left(
 \partial_{\mu} \What ^{a}_{\nu } -
 \partial_{\nu } \What ^{a}_{\mu } + g_{2} \varepsilon ^{abc}
 \What ^{b}_{\mu } \What ^{c}_{\nu } \right)^{2}
- \frac{1}{4} \left(
 \partial_{\mu } \Bhat _{\nu } - \partial_{\nu } \Bhat _{\mu }
 \right)^{2} ,
\eeq
where the isotriplet $\What_{\mu }^{a}$, % and $W_{\mu }^{a}$,
$a=1,2,3$, is associated with the generators $I_{\rw}^{a}$ of the
weak isospin group SU$(2)_{\rw}$ and the isosinglet $\Bhat_{\mu }$
with the weak hypercharge $Y_{\rw}$ of the group U$(1)_{\rY}$.
For later convenience we denote the classical gauge and Higgs
fields with a caret.
The Higgs part has the form
\begin{equation}
{\cal L}_{\mathrm{H}} = \left( \Dhat_{\mu }\Phihat \right) ^{\dagger}
 \left( \Dhat^{\mu }\Phihat \right) -V(\Phihat)
\label{eq:LH}
\end{equation}
with the covariant derivative %in the BFM is given by
\begin{equation} \label{covder}
\Dhat_{\mu } = \partial_{\mu } -i g_{2} I_{\rw}^{a}
\What ^{a}_{\mu }
+ i g_{1} \frac{Y_{\rw}}{2}
\Bhat _{\mu } . %+ B_{\mu }) .
\end{equation}
In \refeq{eq:LH}, $\Phihat(x)$ denotes the complex scalar
SU$(2)_{\rw}$ doublet field of the minimal Higgs sector
with hypercharge $Y_{\rw}^{\Phihat} = 1$
\begin{equation}
\Phihat(x)  = \left( \begin{array}{c} \phihat^{+}(x) \\ \phihat^{0}(x)
\end{array} \right) ,
\end{equation}
and the Higgs potential reads
\beq
V(\Phihat) = \frac{\lambda }{4}
\left(\Phihat^{\dagger}\Phihat\right)^{2}
- \mu ^{2} \Phihat^{\dagger}\Phihat .
\label{eq:Hpot}
\eeq
We write the fermionic part as (neglecting quark mixing as
throughout this paper)
\newcommand{\Ds}{D\hspace{-0.6em}/\hspace{0.1em}}
\begin{eqnarray}
{\cal L}_{\mathrm{F}} &=&
\displaystyle \sum_{k}\,\left(
 \overline{L}_{k}^{\rL} i \hat\Ds {L}_{k}^{\rL}
+\overline{Q}_{k}^{\rL} i \hat\Ds {Q}_{k}^{\rL}
\right)\no%\[1em]
&&\displaystyle \mbox{} + \sum_{k}\,\left(
 \overline{l}_{k}^{\rR} i  \hat\Ds {l}_{k}^{\rR}
+\overline{u}_{k}^{\rR} i  \hat\Ds {u}_{k}^{\rR}
+\overline{d}_{k}^{\rR} i  \hat\Ds {d}_{k}^{\rR}\right) \no
&&\displaystyle \mbox{} - \sum_{k} \left(
 \overline{L}_{k}^{\rL}G^{l}_{k}{l}_{k}^{\rR}\Phihat
+\overline{Q}_{k}^{\rL}G^{u}_{k}{u}_{k}^{\rR}{\hat {\tilde{\Phi}}}
+\overline{Q}_{k}^{\rL}G^{d}_{k}{d}_{k}^{\rR}\Phihat
+ \mathrm{h.c.} \right) .
\label{eq:LF}
\end{eqnarray}
The left-handed fermions of each lepton ($L$) and quark ($Q$)
generation are grouped into SU$(2)_{\rw}$ doublets (the color
index is suppressed)
\beq
{L}_{k}^{\rL}=\omega _{-}  {L}_{k} =
\left( \barr{l} {\nu}_{k}^{\rL} \\ {l}_{k}^{\rL} \earr \right) ,
\qquad
{Q}_{k}^{\rL}=\omega _{-}  {Q}_{k} =
\left( \barr{l} {u}_{k}^{\rL} \\ {d}_{k}^{\rL} \earr \right) ,
\label{eq:fermleft}
\eeq
the right-handed fermions into singlets
\beq
{l}_{k}^{\rR}=\omega _{+}  {l}_{k}, \qquad
{u}_{k}^{\rR}=\omega _{+}  {u}_{k}, \qquad
{d}_{k}^{\rR}=\omega _{+}  {d}_{k},
\label{eq:fermright}
\eeq
where $\omega _{\pm}=(1\pm\gamma _{5})/2$ are the projectors
on right- and left-handed fields, respectively,
$k$ is the generation index, and $\nu$, $l$, $u$ and $d$ stand
for neutrinos, charged leptons, up-type quarks and down-type quarks,
respectively.
The weak hypercharge $Y_{\rw}$ is assigned
according to the Gell-Mann Nishijima relation
\beq
Y_{\rw} = 2 (Q - I^3_{\rw}) ,
\eeq
where $Q$ is the electric charge operator.
In \refeq{eq:LF}, $G_{k}^{l}$, $G_{k}^{u}$ and
$G_{k}^{d}$ denote the Yukawa couplings,
${\hat{\tilde{\Phi}}} =
\left(\phihat^{0*}, -\phihat^{-} \right)^{T}$ is the
charge-conjugated Higgs field, and
$\phihat ^{-}=\left(\phihat^{+}\right)^{*}$.

The physical gauge-boson fields are obtained via
\begin{equation}
\What_{\mu }^{\pm} = \frac{1}{\sqrt{2}} \left( \What_{\mu }^{1} \mp i
\What_{\mu }^{2} \right) , \qquad
\left(\barr{l} \Zhat_{\mu } \\ \Ahat_{\mu } \earr \right) =
\left(\barr{rr} \cw & \sw \\ - \sw & \cw  \earr \right)
\left(\barr{l} \What_{\mu }^{3} \\ \Bhat_{\mu } \earr \right) ,
\label{eq:phyfi}
\end{equation}
where
\beq
\label{eq:cw}
\cw =\cos\theta_{\rw}=\frac{g_{2}}{\sqrt{g_{1}^{2}+g_{2}^{2}}}
     = \frac{\MW}{\MZ} ,
\qquad \sw =\sin\theta_{\rw} = \sqrt{1-\cw^2},
\eeq
and $\theta_{\rw}$ is the weak mixing angle.
The electromagnetic coupling is given by
\beq
\label{eq:e}
e = \frac{g_1g_{2}}{\sqrt{g_{1}^{2}+g_{2}^{2}}};
\eeq
further relations between the physical parameters and the parameters in
$\L_{\mathrm{C}}$ can be found in \citere{Dehab}.

\subsection{Quantization in the background-field method}

In the conventional formalism directly the fields appearing in the
classical Lagrangian are quantized.
A gauge-fixing term is added to ${\cal L}_{\mathrm{C}}$
which breaks the explicit gauge invariance.

Instead, when going from the classical to the quantized theory
in the BFM~\cite{BFMref,Ab81},
the fields $\Vhat$ of ${\cal L}_{\mathrm{C}}$ are
split into classical background fields $\Vhat$ and quantum fields $V$,
\beq
\label{eq:quantLc}
{\cal L}_{\mathrm{C}}(\Vhat) \rightarrow {\cal
L}_{\mathrm{C}}(\Vhat + V) .
\eeq
The quantum fields are the variables of integration in the functional
integral. %of the effective action.
A gauge-fixing term is added which only breaks the
gauge invariance of the quantum fields but retains the gauge
invariance of the effective action with respect to the background
fields. % in the effective action.

In order to avoid tree-level mixing between the gauge bosons and
the corresponding unphysical Higgs bosons, we use
a generalization of the 't Hooft gauge fixing to the BFM~\cite{Sh81}
\beqar\label{tHgf}
\L_{\mathrm{GF}} &=& - \frac{1}{2\xiQ^W}
\biggl[(\de^{ac}\partial_\mu + g_2 \veps^{abc}\hat
W^b_\mu)W^{c,\mu}
       -ig_2\xiQ^W\frac{1}{2}(\hat\Phi^\dagger_i
\si^a_{ij}\Phi_j
                  - \Phi^\dagger_i
\si^a_{ij}\hat\Phi_j)\biggr]^2 \no
                 && {}- \frac{1}{2\xiQ^B}
\biggl[\partial_\mu B^{\mu}
       +ig_1\xiQ^B\frac{1}{2}(\hat\Phi^\dagger_i \Phi_i
                  - \Phi^\dagger_i \hat\Phi_i)\biggr]^2,
\eeqar
where $\si^a$, $a=1,2,3$, denote the Pauli matrices, and $\xiQ^W$,
$\xiQ^B$ are parameters associated with the gauge fixing of the
quantum fields.
The background Higgs field $\hat\Phi$ has the usual non-vanishing
vacuum expectation value $v$, while the one of the quantum Higgs field
$\Phi$ is zero:
\begin{equation}
\hat\Phi(x) = \left( \begin{array}{c}
\phihat ^{+}(x) \\ \frac{1}{\sqrt{2}}\bigl(v + {\hat H}(x) +i
\chihat (x) \bigr)
\end{array} \right) , \qquad
\Phi (x) = \left( \begin{array}{c}
\phi ^{+}(x) \\ \frac{1}{\sqrt{2}}\bigl(H(x) +i \chi (x) \bigr)
\end{array} \right) .
\label{eq:Hbq}
\end{equation}
Here ${\hat H}$ and $H$ denote the physical background and
quantum Higgs field, respectively, and $\phihat ^{+}, \chihat,
\phi ^{+}, \chi$ are unphysical degrees of freedom.
The
gauge-fixing term (\ref{tHgf}) translates to the conventional one upon
replacing the background Higgs field by its vacuum expectation
value and
omitting the background SU$(2)_{\rw}$ triplet field $\hat W^a_\mu$.
Background-field gauge invariance restricts the number of
quantum gauge parameters
to two, one for SU$(2)_{\rw}$ and one for U$(1)_{\rY}$.

In the spirit of the BFM, one should also split the fermion
fields into background and quantum fields.
However, for all fields that do not enter the gauge-fixing term,
quantization in the BFM is equivalent to the
conventional formalism. The Feynman rules for background and
quantum fields are identical for these fields and there is no
need to distinguish them.
We therefore use a common symbol for the fermion fields, i.e.~we
do not write a caret for the fermion background fields.

Next, we express the gauge-fixing term (\ref{tHgf}) by
physical fields. In order to avoid tree-level mixing between the
photon and the \PZ\ boson
one has to chose  $\xiQ = \xiQ^W = \xi_Q^B$. This yields
\beq
\L_{\mathrm{GF}} = - \frac{1}{2 \xiQ}
\left[ (G^{A})^{2} + (G^{Z})^{2} + 2 G^{+} G^{-} \right] ,
\label{eq:LGFph}
\eeq
where
\beqar
G^{A} &=& \partial^{\mu } A_{\mu }
 + i e (\What^+ _{\mu} W^{- \mu} - W^+ _{\mu} \What^{- \mu})
 + i e \xiQ (\phihat^- \phi^+ - \phihat^+ \phi^-) , \no
G^{Z} &=& \partial^{\mu } Z_{\mu }
 - i e \frac{\cw}{\sw} (\What^+ _{\mu} W^{- \mu} - W^+ _{\mu}
  \What^{- \mu})
 - i e \xiQ \frac{\cw^2 - \sw^2}{2 \cw \sw}
  (\phihat^- \phi^+ - \phihat^+ \phi^-) \no
 &&+ e \xiQ \frac{1}{2 \cw \sw}
   (\chihat H - {\hat H} \chi - v \chi) , \no
G^{\pm} &=& \partial^{\mu } W^{\pm}_{\mu }
 \pm i e (\Ahat^{\mu} - \frac{\cw}{\sw} \Zhat^{\mu}) W^{\pm}_{\mu}
 \mp i e (A^{\mu} - \frac{\cw}{\sw} Z^{\mu}) \What^{\pm}_{\mu} \no
&&\mp i e \xiQ \frac{1}{2 \sw}
 \left[ (v + \Hhat \mp i \chihat) \phi^{\pm}
  - (H \mp i \chi) \phihat^{\pm} \right] .
\eeqar

Finally, we add a Faddeev--Popov part to the Lagrangian
\beq
{\cal L}_{\mathrm{FP}} = - \bar{u}^{\alpha } \frac{\delta G^{\alpha
}}{\delta \theta^{\beta }} u^{\beta } ,
\eeq
where $\alpha = A, Z, \pm$, and
$\delta G^{\alpha}/\delta\theta^{\beta}$ is the variation of the
gauge-fixing terms $G^{\alpha }$ under the infinitesimal
quantum gauge transformations of the quantum fields
\beqar
\delta W^{\pm}_{\mu} & = & \partial_{\mu} \delta \theta^{\pm}
 \mp i e (W^{\pm}_{\mu} + \What ^{\pm}_{\mu})
	 (\delta \theta^{A} - \frac{\cw}{\sw} \delta \theta^{Z})
 \pm i e \left[(A_{\mu} + \Ahat_{\mu}) -
      \frac{\cw}{\sw} (Z_{\mu} + \Zhat_{\mu}) \right] \delta\theta^\pm , \no
\delta Z_{\mu} & = & \partial_{\mu} \delta \theta^{Z}
 - i e \frac{\cw}{\sw} \left[ (W^+_{\mu} + \What^+_{\mu})
   \delta \theta^- - (W^-_{\mu} + \What^-_{\mu})
   \delta \theta^+ \right] , \no
\delta A_{\mu} & = & \partial_{\mu} \delta \theta^{A}
 + i e \left[ (W^+_{\mu} + \What^+_{\mu}) \delta \theta^-
  - (W^-_{\mu} + \What^-_{\mu}) \delta \theta^+ \right] , \no
\delta \phi^{\pm} & = & \pm \frac{i e}{2 \sw}
 \left[ H + \Hhat + v \pm i (\chi + \chihat) \right]
 \delta \theta^\pm \mp i e (\phi^\pm + \phihat^\pm)
  (\delta \theta^A - \frac{\cw^2-\sw^2}{2 \cw \sw} \delta
   \theta^Z) , \no
\delta H & = & \frac{i e}{2 \sw} \left[ (\phi^+ + \phihat^+)
 \delta \theta^- - (\phi^- + \phihat^-) \delta \theta^+ \right]
 + \frac{e}{2 \cw \sw} (\chi + \chihat) \delta \theta^Z , \no
\delta \chi & = & \frac{e}{2 \sw} \left[ (\phi^+ + \phihat^+)
 \delta \theta^- + (\phi^- + \phihat^-) \delta \theta^+ \right]
 - \frac{e}{2 \cw \sw} (H + \Hhat + v) \delta \theta^Z .
\label{eq:qugautraf}
\eeqar

Using the Lagrangian specified above, an effective action
$\Gamma[\Vhat, \Shat, F, \bar F]$ is
constructed following \citere{Ab81},
where $\Vhat$ collectively denotes the background gauge fields,
$\Shat$ the background Higgs fields and $F$ the fermion fields.
$\Gamma[\Vhat, \Shat, F, \bar F]$ is
invariant under the background gauge transformations of the background
fields
\beqar
\delta \What^{\pm}_{\mu} & = & \partial_{\mu} \delta \thetahat^{\pm}
 \mp i e \What ^{\pm}_{\mu}
         (\delta \thetahat^{A} - \frac{\cw}{\sw} \delta \thetahat^{Z})
 \pm i e ( \Ahat_{\mu} - \frac{\cw}{\sw} \Zhat_{\mu} )
   \delta \thetahat^{\pm} , \no
\delta \Zhat_{\mu} & = & \partial_{\mu} \delta \thetahat^{Z}
 - i e \frac{\cw}{\sw} ( \What^+_{\mu} \delta \thetahat^-
                        -\What^-_{\mu} \delta \thetahat^+ ) , \no
\delta \Ahat_{\mu} & = & \partial_{\mu} \delta \thetahat^{A}
 + i e ( \What^+_{\mu} \delta \thetahat^-
       - \What^-_{\mu} \delta \thetahat^+ ) , \no
\delta \phihat^{\pm} & = & \pm \frac{i e}{2 \sw}
 ( \Hhat + v \pm i \chihat ) \delta \thetahat^\pm
 \mp i e \phihat^\pm (\delta \thetahat^A
                    - \frac{\cw^2-\sw^2}{2 \cw \sw} \delta
   \thetahat^Z) , \no
\delta \Hhat & = & \frac{i e}{2 \sw} ( \phihat^+ \delta \thetahat^-
                                     - \phihat^- \delta \thetahat^+ )
 + \frac{e}{2 \cw \sw} \chihat \delta \thetahat^Z , \no
\delta \chihat & = & \frac{e}{2 \sw} ( \phihat^+ \delta \thetahat^-
                                     + \phihat^- \delta \thetahat^+ )
 - \frac{e}{2 \cw \sw} (\Hhat + v) \delta \thetahat^Z ,
\label{eq:gautrafo}
\eeqar
and the corresponding transformations of the fermion fields
\beqar
\delta f^{\rL}_u &=&
       \frac{i e}{\sqrt{2} \sw} f^{\rL}_d \delta \thetahat^{+}
      -i e \left[ Q_{f_u} \delta \thetahat^{A} -
    \left(\frac{1}{2\cw\sw} - Q_{f_u} \frac{\sw}{\cw}\right) \delta
\thetahat^{Z}
         \right] f^{\rL}_u , \no
\delta f^{\rL}_d &=& \frac{i e}{\sqrt{2} \sw}
  f^{\rL}_u \delta \thetahat^{-} - i e
  \left[ Q_{f_d} \delta \thetahat^{A} +
  \left(\frac{1}{2\cw\sw} + Q_{f_d} \frac{\sw}{\cw}\right) \delta \thetahat^{Z}
         \right] f^{\rL}_d , \no
\delta f^{\rR} &=& - i e Q_f (\delta \thetahat^{A} +
   \frac{\sw}{\cw} \delta \thetahat^{Z}) f^{\rR} ,
\label{eq:gautraferm}
\eeqar
where $f^{\rL}_u$ stands for all left-handed up-type quarks and
neutrinos of \refeq{eq:fermleft}, $f^{\rL}_d$ denotes their isospin
partners, and $f^{\rR}$ represents the right-handed singlets of
\refeq{eq:fermright}.

The effective action $\Gamma[\Vhat, \Shat, F, \bar F]$ is the generating
functional of the vertex functions which
are obtained by differentiating $\Gamma[\Vhat, \Shat, F, \bar F]$
with respect to its arguments.
The vertex functions can be calculated from Feynman rules that
distinguish between quantum and
background fields. Whereas the quantum fields appear only inside
loops, the background fields are associated with the external lines.
Apart from doubling of the gauge and Higgs fields, the
BFM Feynman rules differ from the conventional ones
only owing to the gauge-fixing and ghost terms, which affect
only vertices that involve both background and quantum fields.
Since the gauge-fixing term is non-linear in the fields, the
gauge parameter enters also the gauge-boson vertices.
As mentioned above, the lowest-order Feynman rules involving
fermion fields are the same as in the conventional formalism.

The S matrix is constructed in the usual way by forming trees
with vertices from $\Gamma[\Vhat, \Shat, F, \bar F]$
which are connected by
lowest-order background-field propagators~\cite{Ab83}.
As a simple example, we calculated the one-loop process
$\PZ \rightarrow \Pb {\bar\Pb}$ for arbitrary values of $\xiQ$.
We verified that the resulting S-matrix element is in fact independent
of $\xiQ$ and equal to the one obtained in the conventional formalism.

We have evaluated the complete set of BFM Feynman rules in the
electroweak SM for arbitrary values
of the quantum gauge parameter $\xiQ$.
They are listed in the appendix.
Despite the distinction between background
and quantum fields, calculations in the BFM become in general simpler
than in the conventional formalism.
This is in
particular the case in the 't Hooft-Feynman gauge ($\xiQ = 1$)
for the quantum fields
where many vertices simplify considerably (see appendix).
Moreover, the gauge
fixing of the background fields %,
is totally unrelated
to the gauge fixing of the quantum fields. This freedom can be
used to choose a particularly suitable background gauge,
e.g.~the unitary gauge or a non-linear gauge~\cite{Ga81}.
In this way the number of Feynman diagrams can be
reduced drastically.
The gauge fixing of the background fields does not affect
$\Gamma[\Vhat, \Shat, F, \bar F]$.
It is only relevant for the construction of connected Green
functions and S-matrix elements. In particular, in linear
background gauges only the tree-level propagators are concerned.

Since the background gauge parameters enter only tree-level
quantities, their cancellation in S-matrix elements is a direct
consequence of the
BFM \WI. As an example, this can easily be checked
for background $R_{\xi}$ gauges in four-fermion processes.
In this case, the BFM \WI\ imply the cancellation of the background
gauge parameters
separately for \se\ and vertex contributions.

In \citere{Be93}, the BFM was applied to the process
$\PZ \rightarrow 3 \ga$ at one-loop order.
However, the gauge-fixing term used there breaks
background-field gauge invariance since no background Higgs field
has been introduced. This influences the vertex functions with
external Higgs fields. Since for the specific process
treated in~\citere{Be93} no such vertex function contributes, the
results obtained there are nevertheless unaffected. In~\citere{Be93},
the Feynman rules  for vertices involving exactly two quantum fields and
no background Higgs fields
were given for the special case
$\xiQ = 1$. Putting $\xiQ = 1$ in the corresponding
rules given in appendix~\ref{app:Frules} we find agreement
except for the ones in \refeq{fr:Cvvgg} which differ by a factor 2.

\section{WARD IDENTITIES}
\label{sec:WI}

The invariance of the effective action under the
background gauge transformations specified
in \refeq{eq:gautrafo} and \refeq{eq:gautraferm},
\beqar
\de\Ga &=&
\sum_{i} \frac{\delta \Gamma[\Vhat, \Shat, F, \bar F]}{\delta \Vhat_i}
\delta \Vhat_i +
\sum_{j} \frac{\delta \Gamma[\Vhat, \Shat, F, \bar F]}{\delta \Shat_j}
\delta \Shat_j \no
&& {} + \sum_{k} \left( \delta \bar F_k
\frac{\delta \Gamma[\Vhat, \Shat, F, \bar F]}{\delta \bar F_k} -
\frac{\delta \Gamma[\Vhat, \Shat, F, \bar F]}{\delta F_k}
\delta F_k \right) = 0 ,
\label{eq:inv}
\eeqar
where $i,j,k$ run over all background gauge fields, background
Higgs fields and fermion fields, respectively,
gives rise to simple Ward identities.
Since the gauge invariance has been retained in the
background-field formulation, these are precisely the Ward
identities related to the classical Lagrangian.
This is in contrast to the conventional formalism where owing to
the gauge-fixing procedure the explicit gauge invariance is lost
and the Ward identities are obtained from the invariance
under BRS transformations.
These Slavnov--Taylor identities have a more complicated
structure and in general involve ghost contributions
(see e.g.~\citere{BHS}).

The BFM Ward identities follow from
differentiating~(\ref{eq:inv}) with respect to the fields
and are valid in all orders of perturbation theory. Note that
the identities hold for arbitrary values of the quantum gauge
parameter.

We first list some identities for \ses. These are related to the
two-point vertex functions as follows
\beqar
\Ga^{\Vhat \Vhat'}_{\mu \nu} (k, -k) &=&
i (- g_{\mu \nu} k^2 + k_{\mu} k_{\nu} + g_{\mu \nu}
    M_{\Vhat}^2) \delta_{\Vhat \Vhat'} \no
&& \mbox{} + i \left(- g_{\mu \nu}
    + \frac{k_{\mu} k_{\nu}}{k^2} \right) \Si^{\Vhat \Vhat'}_{\rT}(k^2)
    - i \frac{k_{\mu} k_{\nu}}{k^2} \Si^{\Vhat \Vhat'}_{\rL}(k^2) ,
 \label{eq:vertsemunu}    \no
\Ga^{\What^{\pm} {\phihat^{\mp}}}_{\mu} (k, -k) &=& i k_{\mu}
    \left[\pm\MW + \Si^{\What^{\pm} {\phihat^{\mp}}}(k^2) \right] , \no
\Ga^{\Zhat \chihat}_{\mu} (k, -k) &=& i k_{\mu} \left[i\MZ +
    \Si^{\Zhat\chihat}(k^2)\right] , \no
\Ga^{\Ahat\chihat}_{\mu} (k, -k) &=& i k_\mu \Si^{\Ahat\chihat}(k^2) , \no
\Ga^{\Ahat\Hhat}_{\mu} (k, -k) &=& i k_\mu \Si^{\Ahat\Hhat}(k^2) , \no
\Ga^{\Hhat\Hhat} (k, -k) &=& i (k^2 - \MH^2)
    + i \Si^{\Hhat\Hhat}(k^2) , \no
\Ga^{\chihat \chihat} (k, -k) &=& i k^2
    + i \Si^{\chihat \chihat}(k^2) , \no
\Ga^{\phihat^{+} \phihat^{-}} (k, -k) &=& i k^2
    + i \Si^{\phihat^{+} \phihat^{-}}(k^2) , \no
\Ga^{\Hhat} &=& i T^{\Hhat} , \no
\Ga^{\bar ff}(p,-p) &=& -i(\ps + m_f) - i\ps\om_-\Si^{\bar ff}_\rL(p^2)
    - i\ps\om_+\Si^{\bar ff}_\rR(p^2) + im_f\Si^{\bar ff}_\rS(p^2)
\eeqar
where $\Vhat, \Vhat'$ indicate vector fields,
and, as throughout this paper, all momenta and fields in
the vertex functions are incoming.
In the following we omit the second argument of the two-point vertex
functions which is fixed by momentum conservation.
Note that
no gauge-fixing terms for the background fields are included in
the vertex functions,
i.e.~the lowest-order
contributions to the vertex functions follow directly from
${\cal L}_{\mathrm{C}}$.
The \ses\ contain no tadpole contributions;
these appear explicitly as $T^{\Hhat}$. We
obtain the following \WI\ for the \ses\
\beqar
\label{eq:sega}
\Si^{\Ahat\Ahat}_\rL(k^2) &=& 0, \\
\label{eq:segaZ}
\Si^{\Ahat\Zhat}_\rL(k^2) &=& 0, \\
\label{eq:segachi}
\Si^{\Ahat\hat\chi}(k^2)  &=& 0, \\
\label{eq:segaH}
\Si^{\Ahat\Hhat}(k^2)  &=& 0, \\
\label{eq:seZ1}
\Si^{\Zhat\Zhat}_\rL(k^2) - i \MZ \Si^{\Zhat\hat\chi}(k^2)
&=& 0, \\
\label{eq:seZ2}
k^2 \Si^{\Zhat\hat\chi}(k^2) - i \MZ
\Si^{\hat\chi\hat\chi}(k^2) + i \frac{e}{2 \cw \sw} T^{\Hhat} &=& 0 , \\
\label{eq:seW1}
\Si^{\What^{\pm}\What^{\mp}}_\rL(k^2) \mp \MW
\Si^{\What^{\mp}\phihat^{\pm}}(k^2) &=& 0, \\
\label{eq:seW2}
k^2 \Si^{\What^{\pm}\phihat^{\mp}} \mp \MW
\Si^{\phihat^{\pm}\phihat^{\mp}} \pm \frac{e}{2 \sw} T^{\Hhat}
&=& 0 .
\eeqar

As a direct consequence of the Ward identities \refeq{eq:sega} and
\refeq{eq:segaZ} and of the analyticity of
$\Ga^{\Ahat\Ahat}_{\mu \nu}(k)$ and
$\Ga^{\Ahat\Zhat}_{\mu \nu}(k)$ at $k^2 = 0$
their transverse parts vanish at zero momentum, i.e.
\beq
\Si^{\Ahat\Ahat}_\rT(0) = 0 ,
\label{eq:ga0}
\eeq
and
\beq
\Si^{\Ahat\Zhat}_\rT(0) = 0 .
\label{eq:gaz0}
\eeq
Whereas the QED relations \refeqs{eq:sega}
and \refeqf{eq:ga0} %and \refeq{eq:segaH}
are valid in the BFM to all orders, they only hold at
one-loop order in the conventional formalism.
The identities (\ref{eq:segaZ}), (\ref{eq:segachi}) and
\refeq{eq:gaz0} have no conventional counterpart.
Note that the vanishing of the photon--\PZ-boson mixing at
zero momentum is explicitly enforced through a renormalization
condition in the usual on-shell scheme (see e.g.~\citere{Dehab}),
while in the BFM it is automatically fulfilled as a
consequence of gauge invariance.
The impact of the BFM on the renormalization program will be
discussed in more detail in the next section.
Equation \refeq{eq:segachi} shows
that in contrast to the $R_{\xi}$ gauges of the conventional
formalism the photon does not mix with the unphysical
scalar $\chi$ in the BFM. For the \PZ-boson
\se\ one has in the usual formalism at one-loop
order\footnote{For the relations corresponding to
(\ref{eq:sega}) and \refeq{eq:sezconv} in the conventional
formalism at two-loop order see~\citere{2loop}.}
\beq
\label{eq:sezconv}
k^2\Bigl[\Si^{\FZ\FZ}_\rL(k^2) - 2i \MZ \Si^{\FZ\chi}(k^2)\Bigr]
 -  \MZ^2 \Si^{\chi\chi}(k^2) +  \frac{e\MZ}{2 \cw \sw} T^\FH = 0.
\eeq
In the BFM, this relation decouples
into two simpler Ward identities, (\ref{eq:seZ1}) and
(\ref{eq:seZ2}), which are valid to all orders.

The three-point function $\Ga^{\Ahat\Ffbar\Ff}_{\mu}$ obeys
\beq
\label{WIAff}
k^\mu \Ga^{\Ahat\Ffbar\Ff}_{\mu}(k,\bar p, p) = -e\Qf
[\Ga^{\Ffbar\Ff}(\bar p) - \Ga^{\Ffbar\Ff}(-p)],
\eeq
i.e.~just the QED Ward identity.
Note that despite the U$(1)_{\mathrm{em}}$ gauge
invariance of the classical Lagrangian the conventional
formalism does not yield the QED-type Ward identity
in the SM.
In the BFM, the Ward identities for the $\Zhat\Ffbar\Ff$ and
$\What\Ffbar\Ff$ vertices read
\beqar
\label{WIZff}
k^\mu \Ga^{\Zhat\Ffbar\Ff}_{\mu}(k,\bar p,p) -i
\MZ \Ga^{\hat\chi\Ffbar\Ff}(k,\bar p,p) &=& e
[\Ga^{\Ffbar\Ff}(\bar p)(\vf-\af\ga_5) - (\vf+\af\ga_5)
\Ga^{\Ffbar\Ff}(-p)], \no
&& \\
\label{WIWff}
k^\mu \Ga^{\What^+\Ffbar_u\Ff_d}_{\mu}(k,\bar p,p) -
\MW \Ga^{\phihat^+\Ffbar_u\Ff_d}(k,\bar p,p) &=&
\frac{e}{\sqrt{2}\sw}
[\Ga^{\Ffbar_u\Ff_u}(\bar p)\om_- - \om_+ \Ga^{\Ffbar_d\Ff_d}(-p)] , \no
k^\mu \Ga^{\What^-\Ffbar_d\Ff_u}_{\mu}(k,\bar p,p) +
\MW \Ga^{\phihat^-\Ffbar_d\Ff_u}(k,\bar p,p) &=&
\frac{e}{\sqrt{2}\sw}
[\Ga^{\Ffbar_d\Ff_d}(\bar p)\om_- - \om_+ \Ga^{\Ffbar_u\Ff_u}(-p)] ,
\eeqar
where $\vf = (I^3_{\rw,f} - 2\sw^2 Q_f)/(2\sw\cw)$
and $\af = I^3_{\rw,f}/(2\sw\cw)$.
Also the triple
gauge-boson vertex fulfills a QED-like Ward identity
\beq
\label{WIAWW}
k^\mu \Ga^{\Ahat\What^+\What^-}_{\mu\rho\si}(k,k_+,k_-) =
e [\Ga^{\What^+\What^-}_{\rho\si}(k_+) -
\Ga^{\What^+\What^-}_{\rho\si }(-k_-)] .
\eeq
An example involving vertex functions with Higgs bosons is:
\beq
\label{WIZXH}
k^\mu_\FZ \Ga^{\Zhat\hat\chi\Hhat}_{\mu}(k_\FZ,k_\chi,k_\FH)
-i\MZ \Ga^{\hat\chi\hat\chi\Hhat}(k_\FZ,k_\chi,k_\FH) =
-i\frac{e}{2\sw\cw} [\Ga^{\Hhat\Hhat}(k_\FH) -
\Ga^{\hat\chi\hat\chi}(k_\chi)] .
\eeq
{}Further \WI\ are listed in~\citeres{bgf,bgfproc}.

\section{RENORMALIZATION OF THE STANDARD MODEL}
\label{se:renco}

As we will show in this section, the BFM gauge invariance has
important consequences for the structure of the
renormalization constants necessary to render Green functions
and S-matrix elements finite.
The arguments which we give in the following are made
explicit for the one-loop level.
It is easy, however, to extend them by induction to arbitrary
orders in perturbation theory.

Following the QCD treatment of~\citere{Ab81},
we introduce field renormalization only for the background fields.
We start with the following set of renormalization constants
for the parameters
\beqar
e_0 &=& Z_e e = (1 + \delta Z_e) e , \no
{\MW^2}_{,0} &=& \MW^2 + \delta \MW^2 , \no
{\MZ^2}_{,0} &=& \MZ^2 + \delta \MZ^2 , \no
{\MH^2}_{,0} &=& \MH^2 + \delta \MH^2 , \no
{\Mf}_{,0} &=& \Mf + \delta \Mf , \no
t_0 &=& t + \delta t ,
\label{eq:renconsts1}
\eeqar
and fields
\beqar
\What_{0}^{\pm}  & = & Z_{\What}^{1/2} \What^{\pm}
  = (1+\frac{1}{2}\delta Z_{\What}) \What^{\pm} , \no
\left(\barr{l} \Zhat_{0} \\ \Ahat_{0} \earr \right)  & = &
\left(\barr{ll} Z_{\Zhat\Zhat}^{1/2} & Z_{\Zhat\Ahat}^{1/2}  \\[1ex]
                Z_{\Ahat\Zhat}^{1/2} & Z_{\Ahat\Ahat}^{1/2}
      \earr
\right)
\left(\barr{l} \Zhat \\ \Ahat \earr \right)   =
\left(\barr{cc} 1 + \frac{1}{2}\delta Z_{\Zhat\Zhat} &
\frac{1}{2}\delta Z_{\Zhat\Ahat} \\ [1ex]
\frac{1}{2}\delta Z_{\Ahat\Zhat}  & 1 + \frac{1}{2}\delta
Z_{\Ahat\Ahat}
\earr \right)
\left(\barr{l} \Zhat \\[1ex] \Ahat \earr \right)  , \no
\Hhat_{0} & = & Z_{\Hhat}^{1/2} \Hhat
 = (1+\frac{1}{2}\delta Z_{\Hhat}) \Hhat, \no
\chihat_{0} & = & Z_{\chihat}^{1/2} \chihat =
  (1+\frac{1}{2}\delta Z_{\chihat}) \chihat , \no
\phihat_{0}^{\pm} & = & Z_{\phihat}^{1/2} \phihat^{\pm} =
  (1+\frac{1}{2}\delta Z_{\phihat}) \phihat^{\pm} , \no
f_{0}^{\rL} & = & \left(Z^{\rL}_{f}\right)^{1/2} f^{\rL}
 =(1 + \frac{1}{2}\delta Z^{\rL}_{f})  f^{\rL} , \no
f_{0}^{\rR} & = & \left(Z^{\rR}_{f}\right)^{1/2} f^{\rR}
 =(1 + \frac{1}{2}\delta Z^{\rR}_{f})  f^{\rR} .
\label{eq:renconsts2}
\eeqar
The tadpole counterterm $\delta t$ renormalizes the term
in the Lagrangian linear
in the Higgs field $\Hhat$ which we denote by $t \Hhat(x)$ with
$t = v (\mu^2 - \la v^2/4)$.
It corrects for the shift in the minimum of the Higgs potential
due to radiative corrections. Choosing $v$ as the correct vacuum
expectation value of the Higgs field $\Phihat$ is equivalent to
the vanishing of $t$.

In order to preserve the background-field gauge invariance when
renormalizing the theory it is necessary to require that the
renormalized vertex functions fulfill Ward identities of the same form
as the unrenormalized ones.
As a consequence, also the counterterms have to fulfill these Ward
identities. %, in particular \refeqs{eq:sega} -- \refeqf{WIZXH}.
This yields relations between the counterterms.

{}For example, from \refeq{eq:segaZ} one obtains immediately
\beq
0=\Si^{\Ahat\Zhat,\ren}_\rL(k^2) =
\Si^{\Ahat\Zhat}_\rL(k^2) - \MZ^2 \frac{1}{2} \delta Z_{\Zhat\Ahat} =
- \MZ^2 \frac{1}{2} \delta Z_{\Zhat\Ahat} ,
\eeq
\ie
\beq \label{eq:delZZA}
\delta Z_{\Zhat\Ahat} = 0.
\eeq
Expressing bare quantities in the QED Ward identity \refeq{WIAff}
through renormalized ones and counterterms yields
\bma
k^\mu \Ga^{\Ahat\Ffbar\Ff}_{\mu}(k,\bar p, p) =
k^\mu \Ga^{\Ahat\Ffbar\Ff, \ren}_{\mu}(k,\bar p, p) + i e Q_f \ks
\left(\delta Z_e + \frac{1}{2} \delta Z_{\Ahat\Ahat} +
\delta Z^{\rR}_{e} \omega _{+} + \delta Z^{\rL}_{e} \omega _{-}
\right) ,
\ema
where \refeq{eq:delZZA} was used, and
\bma
-eQ_f \Bigl[\Ga^{\Ffbar\Ff}(\bar p) - \Ga^{\Ffbar\Ff}(-p)] =
-eQ_f \Bigl[\Ga^{\Ffbar\Ff, \ren}(\bar p) - \Ga^{\Ffbar\Ff, \ren}(-p)]
+ i e Q_f \ks \left(\delta Z^{\rR}_{e} \omega _{+} +
\delta Z^{\rL}_{e} \omega _{-} \right) .
\ema
Using the Ward identity both for bare and renormalized
quantities implies
\beq
\delta Z_{\Ahat\Ahat} = - 2 \delta Z_e.
\label{eq:delZaa}
\eeq
This is just the famous
relation between the renormalizations of field and coupling
known from QED.
In contrast to the conventional formalism, the BFM yields
this relation also for the electroweak SM.
Note that after fixing the charge renormalization there is no
more freedom to impose an extra condition for the field renormalization.
Just as in QED, %also in the SM in the BFM
the on-shell definition of the electric
charge together with gauge invariance automatically fixes the residue
of the photon propagator to unity. This
can be derived using %from the renormalization condition
the Ward identities \refeqs{eq:gaz0} and \refeq{WIAff}.
Instead of considering~\refeq{WIAff}, the relation \refeq{eq:delZaa}
can equivalently be
obtained from the Ward identity~\refeq{WIAWW} for
the non-Abelian coupling.

{}From the Ward identities
\refeq{WIZff}, \refeq{WIWff} and \refeq{WIZXH} one
derives in a similiar way the following relations between the
renormalization constants
\beqar
\label{eq:delZB}
\delta Z_{\Ahat\Zhat} &=& 2 \frac{\cw}{\sw}
    \frac{\delta \cw ^2}{\cw ^2} , \no
\delta Z_{\Zhat\Zhat} &=& - 2 \delta Z_e -
    \frac{\cw ^2 - \sw ^2}{\sw^2} \frac{\delta \cw ^2}{\cw ^2} , \no
\delta Z_{\What} &=& - 2 \delta Z_e -
    \frac{\cw ^2}{\sw^2} \frac{\delta \cw ^2}{\cw ^2} , \no
\delta Z_{\Hhat} &=& \delta Z_{\chihat} = \delta Z_{\phihat} %\no &=&
      = - 2 \delta Z_e -
	\frac{\cw ^2}{\sw^2} \frac{\delta \cw ^2}{\cw ^2} +
	\frac{\delta \MW^2}{\MW^2} ,
\eeqar
where
\bma
\frac{\delta \cw ^2}{\cw ^2} =
\frac{\delta \MW^2}{\MW^2} - \frac{\delta \MZ^2}{\MZ^2} .
\ema
Finally, we get for the field renormalizations of the fermions
\beq
\delta Z^{\rL}_{f} = \delta Z^{\rL}_{f_u} = \delta Z^{\rL}_{f_d} ,
\label{eq:zferm}
\eeq
i.e.~the field renormalization constants for the two left-handed
fermions in a doublet must be equal.

The relations \refeq{eq:delZZA} -- \refeqf{eq:delZB} express the
field renormalization constants of all gauge bosons and scalars
completely in terms of the renormalization constants of the
electric charge and the particle masses. If the renormalized
parameters are identified with the physical electron charge and
the physical particle masses, they are manifestly
gauge-independent. Moreover, the bare quantities $g_{1,0}$,
$g_{2,0}$, $\lambda_0$, $\mu_0$ and $G^f_{i,0}$ in the
Lagrangian obviously are
also gauge-independent, as they represent free parameters of the
theory. According to \refeq{eq:cw} and \refeq{eq:e}, the same is
true for the bare charge
and the bare weak mixing angle. Consequently, the counterterms
$\delta Z_e$ and $\delta\cw^2$ for the gauge couplings are
gauge-independent. The relations \refeq{eq:delZaa} and
\refeq{eq:delZB} therefore imply that the field renormalizations
of all gauge-boson fields are gauge-independent. This is in
contrast to the conventional formalism where the field
renormalizations in the on-shell scheme are gauge-dependent.

It should be recalled at this point that in contrast to
$\delta Z_e$ and $\delta\cw^2$ the counterterms for the masses
are not gauge-independent. This can be traced back to the mechanism
of spontaneous symmetry breaking. The non-vanishing
vacuum expectation value of the Higgs field, which generates the mass
terms, is clearly not invariant under gauge transformations.
Whereas the renormalized value $v=2\sw\MW/e$ is
gauge-independent, the bare quantity $v_0$ and the corresponding
counterterm $\delta v$ are not~\cite{Lee}.
As a consequence, the bare masses which depend on $v_0$ are
gauge-dependent.
Thus, the counterterms
$\delta\MW^2$, $\delta\MZ^2$, $\delta\MH^2$, $\delta\Mf$ and
$\delta t$ are also
gauge-dependent. The physical
masses, however, are determined by the pole positions of the
propagators, i.e.~the zeros of
$k^2 - M^2 - \delta M^2 + C \delta t/\MH^2 + \Sigma(k^2) +
C T^{\Hhat}/\MH^2$,
where $C$ denotes the coupling of the fields to the Higgs field.
The linear combination $\delta M^2 - C \delta t/\MH^2$ of the
mass and tadpole counterterm is
independent of $\delta v$ and thus gauge-independent.

The relations \refeq{eq:delZZA} -- \refeqf{eq:delZB}
reduce the number of independent
renormalization constants considerably. One is left with the
parameter renormalizations appearing in~\refeqs{eq:renconsts1}
and the fermion field renormalization constants $\delta Z^{\rL}_{f}$,
$\delta Z^{\rR}_{f_u}$ and $\delta Z^{\rR}_{f_d}$.
We choose on-shell renormalization
conditions for the parameters as in~\citere{Dehab}\footnote{The
charge renormalization condition formulated in~\citere{Dehab}
assumes that the residue of the renormalized photon propagator equals
unity and that the photon--\PZ-boson mixing vanishes for on-shell
photons.
Owing to the \WI, these conditions are fulfilled and we
can use the same condition in the BFM.} and express the
renormalization constants in terms of unrenormalized \ses\
and the tadpole
\beqar
\label{eq:rencond}
\delta Z_e &=& \frac{1}{2} \frac{\partial
\Si^{\Ahat\Ahat}_{\rT}(k^2)}{\partial k^2} \Biggl|_{k^2 = 0}
, \no
\delta \MW^2 &=& \Re\left(\Si^{\What\What}_{\rT}(\MW^2)\right),\no
	 %- \frac{e \MW}{\sw \MH^2} T^{\Hhat} \Bigr],\no
\delta \MZ^2 &=& \Re\left(\Si^{\Zhat\Zhat}_{\rT}(\MZ^2)\right),\no
	 %- \frac{e \MW}{\cw^2 \sw \MH^2} T^{\Hhat} \Bigr],\no
\delta \MH^2 &=& \Re\left(\Si^{\Hhat\Hhat}_{\rT}(\MH^2)\right),\no
	 %- \frac{e \MW}{\sw \MH^2} T^{\Hhat} \Bigr],\no
\delta \Mf &=& \frac{1}{2}\Mf \Re\Bigr[\Si^{\Ffbar f}_{\rL}(\Mf^2)
                                     + \Si^{\Ffbar f}_{\rR}(\Mf^2)
                                     + 2\Si^{\Ffbar f}_{\rS}(\Mf^2)
                 \Bigl] ,\no
\delta t &=& - T^{\Hhat} .
\eeqar
The fermion field renormalization constants can be fixed as follows
\beqar
\label{eq:rencondferm}
\delta Z_f^{\rL} &=& - \Re\Si^{\bar f_df_d}_{\rL}(m_{f_d}^2) -
 m_{f_d}^2 \frac{\partial}{\partial k^2}
 \Re \left(\Si^{\bar f_df_d}_{\rL}(k^2) + \Si^{\bar f_df_d}_{\rR}(k^2)
 + 2 \Si^{\bar f_df_d}_{\rS}(k^2) \right) \biggr|_{k^2 = m_{f_d}^2} ,\no
\delta Z_{f_{u}}^{\rR} &=&
 - \Re\Si^{\bar f_uf_u}_{\rR}(m_{f_{u}}^2) -
 m_{f_{u}}^2 \frac{\partial}{\partial k^2}
 \Re \left(\Si^{\bar f_uf_u}_{\rL}(k^2) + \Si^{\bar f_uf_u}_{\rR}(k^2)
 + 2 \Si^{\bar f_uf_u}_{\rS}(k^2) \right)
 \biggr|_{k^2 = m_{f_{u}}^2}  ,\no
\delta Z_{f_{d}}^{\rR} &=&
 - \Re\Si^{\bar f_df_d}_{\rR}(m_{f_{d}}^2) -
 m_{f_{d}}^2 \frac{\partial}{\partial k^2}
 \Re \left(\Si^{\bar f_df_d}_{\rL}(k^2) + \Si^{\bar f_df_d}_{\rR}(k^2)
 + 2 \Si^{\bar f_df_d}_{\rS}(k^2) \right)
\biggr|_{k^2 = m_{f_{d}}^2} .
\eeqar

Although there is no freedom to choose the field renormalizations
of the gauge bosons, scalars and left-handed up-type fermions
in the BFM, the specified
set of renormalization constants is still sufficient to render
all background-field vertex functions finite\footnote{Beyond one-loop
order one needs in addition a renormalization of
the quantum gauge parameters \cite{Ab81}. At the one-loop
level these counterterms
do not enter the background-field vertex functions because
$\xi_Q$ does not appear in pure background-field vertices.
Clearly, the renormalization of gauge parameters is irrelevant
for S-matrix
elements at any order as these are gauge-independent.}.
This is evident since the divergences of the vertex functions are
subject to the same restriction as the counterterms.
In order to illustrate this fact at one-loop order we list the
divergent part of the \ses\ in the BFM using dimensional
regularization and writing the dimension as $D = 4 - \eps$,
\def\api{\frac{e^2}{16\pi^2}}
\beqar
\label{eq:ses}
\left(\Si^{\Ahat\Ahat}_\rT(k^2)\right)^{\dive} &=&
\api  k^2 \left(\frac{32}{9}n - 7 \right) \frac{2}{\eps} \; , \no
\left(\Si^{\Ahat\Zhat}_\rT(k^2)\right)^{\dive} &=&
\api  k^2 \left(\frac{32\sw^2 - 12}{9\cw\sw}n
+ \frac{42\cw^2+1}{6\cw\sw}\right) \frac{2}{\eps} \;,\no
\left(\Si^{\Zhat\Zhat}_\rT(k^2)
\right)^{\dive} &=&
\api \Biggl[k^2 \left(\frac{32\sw^4 - 12(2\sw^2 - 1) }{9\cw^2\sw^2}n
- \frac{42\cw^4+2\cw^2-1}{6\cw^2\sw^2}\right) \no
&& \mbox{} - \sum_{f}\frac{\Mf^2}{2 \cw^2\sw^2} +
\frac{2 \MW^2 + \MZ^2}{4 \cw^2\sw^2} (\xiQ + 3)\Biggr]
\frac{2}{\eps} \;,\no
\left(\Si^{\What\What}_\rT(k^2)
\right)^{\dive} &=&
\api \Biggl[k^2 \left(\frac{4}{3\sw^2}n - \frac{43}{6\sw^2}\right)
- \sum_{f}\frac{\Mf^2}{2 \sw^2} + \frac{2 \MW^2 + \MZ^2}{4
\sw^2} (\xiQ + 3)\Biggr]
\frac{2}{\eps} \;,\no
\left(\Si^{\Hhat\Hhat}(k^2)\right)^{\dive} &=&
\api \Biggl[k^2 \left(\sum_{f}\frac{\Mf^2}{2 \MW^2 \sw^2}
- \frac{2 \cw^2 + 1}{4 \cw^2\sw^2} (\xiQ + 3)\right)
- \sum_{f}\frac{3 \Mf^4}{\MW^2 \sw^2} \no
&& {} + \frac{3 (5 \MH^4 + 12 \MW^4 + 6 \MZ^4) +
5 \MH^2 (2 \MW^2 + \MZ^2) \xiQ}{8 \MW^2 \sw^2}
\Biggr] \frac{2}{\eps} \;,\no
\left(\Si^{\chihat\chihat}(k^2)\right)^{\dive} &=&
\left(\Si^{\phihat\phihat}(k^2)\right)^{\dive} \no
&=&
\api \Biggl[k^2 \left(\sum_{f}\frac{\Mf^2}{2 \MW^2 \sw^2}
- \frac{2 \cw^2 + 1}{4 \cw^2\sw^2} (\xiQ + 3)\right)
\Biggr] \frac{2}{\eps}
+\frac{e}{2\MW\sw} \left(T^{\Hhat}\right)^{\dive} , \no
\left(\Si^{\bar ff}_\rL(k^2)\right)^{\dive} &=&
\api \left(\frac{\Mf^2 + m_{\Pf'}^2}{4 \MW^2 \sw^2} +
\frac{4 \sw^2 Q_f^2 - 8 I^3_{\rw,f} Q_f \sw^2 +
2 \cw^2 + 1}{4 \cw^2\sw^2} \xiQ\right) \frac{2}{\eps} \;,\no
\left(\Si^{\bar ff}_\rR(k^2)\right)^{\dive} &=&
\api \left(\frac{\Mf^2}{2 \MW^2 \sw^2} +
\frac{Q_f^2}{\cw^2} \xiQ\right) \frac{2}{\eps} \;,\no
\left(\Si^{\bar ff}_\rS(k^2)\right)^{\dive} &=&
- \api \left(\frac{m_{\Pf'}^2}{2 \MW^2 \sw^2} +
\frac{(Q_f - I^3_{\rw,f}) Q_f}{\cw^2}(\xiQ + 3)\right)
\frac{2}{\eps} \;,\no
\left(T^{\Hhat}\right)^{\dive} &=& \api \Biggl(
- \sum_\Pf\frac{2 \Mf^4}{e \MW \sw} \no
&& {} + \frac{3 (\MH^4 + 4 \MW^4 + 2 \MZ^4) +
\MH^2 (2 \MW^2 + \MZ^2) \xiQ}{4 e \MW \sw} \Biggr)
\frac{2}{\eps} \;,
\eeqar
where $f'$ is the isospin partner of fermion $f$,
$n$ denotes the number of fermion generations and
the summations run over all fermion flavors and colors.
The fermion \ses\ and the fermionic contributions to the
gauge-boson and scalar \ses\ are included for completeness.
They have the same form as in the conventional formalism.

Using \refeqs{eq:rencond} and \refeqs{eq:rencondferm}
we obtain the divergent parts of the
renormalization constants
\beqar
\left(\delta Z_e \right)^{\dive} &=&
  \api \frac{1}{2} \left(\frac{32}{9}n - 7 \right) \frac{2}{\eps} \;,\no
\left(\frac{\delta \MW^2}{\MW^2}\right)^{\dive} &=&
  \api \left(\frac{4}{3\sw^2}n -
  \frac{43}{6\sw^2} - \sum_{f}\frac{\Mf^2}{2 \MW^2 \sw^2} +
  \frac{2 \cw^2 + 1}{4 \cw^2\sw^2} (\xiQ + 3)\right)
  \frac{2}{\eps} \;,\no
\left(\frac{\delta \cw^2}{\cw^2}\right)^{\dive} &=& \api \left(
  \frac{- 32 \sw^2 + 12}{9 \cw^2} n - 7 - \frac{1}{6 \cw^2}
  \right) \frac{2}{\eps} \;,\no
\left(\delta \MH^2\right)^{\dive} &=& \api \biggl(
  \sum_{f}\frac{\Mf^2 (\MH^2 - 6 \Mf^2)}{2 \MW^2 \sw^2} \no
  && {} +  \frac{3[5 \MH^4 + 12 \MW^4 + 6 \MZ^4 +
  \MH^2 (2 \MW^2 + \MZ^2) (\xiQ - 2)]}{8\MW^2 \sw^2}
  \biggr) \frac{2}{\eps} \;,\no
\left(\delta \Mf\right)^{\dive} &=& \api \Mf
  \biggr( \frac{3 (\Mf^2 - m_{\Pf'}^2)}{8 \MW^2 \sw^2} -
  \frac{3 Q_f(Q_f-I_{\rw,3}^f)}{\cw^2}
  + \frac{(2\cw^2 + 1)\xiQ}{8 \cw^2\sw^2}
  \biggl) \frac{2}{\eps} \;,\no
\left(\delta t\right)^{\dive} &=& - \left(T^{\Hhat}\right)^{\dive} ,\no
\left(\delta Z_f^{\rL}\right)^{\dive} &=&
  - \left(\Si^{\bar f_uf_u}_{\rL}(k^2)\right)^{\dive} =
  - \left(\Si^{\bar f_df_d}_{\rL}(k^2)\right)^{\dive} , \no
\left(\delta Z_{f_{u}}^{\rR}\right)^{\dive} &=&
  - \left(\Si^{\bar f_uf_u}_{\rR}(k^2)\right)^{\dive} , \no
\left(\delta Z_{f_{d}}^{\rR}\right)^{\dive} &=&
  - \left(\Si^{\bar f_df_d}_{\rR}(k^2)\right)^{\dive} .
\eeqar
According to \refeq{eq:delZaa} and \refeqs{eq:delZB}, this also
fixes the divergent parts of the gauge-boson and scalar field
renormalization constants yielding
\beqar
\label{eq:fieldrc}
\left(\delta Z_{\Ahat\Ahat} \right)^{\dive} &=&
  - \api \left(\frac{32}{9}n - 7 \right) \frac{2}{\eps} \;,\no
\left(\delta Z_{\Ahat\Zhat} \right)^{\dive} &=&
  - \api 2 \left(\frac{32 \sw^2 - 12}{9 \cw \sw} n +
  \frac{42 \cw^2 + 1}{6 \cw\sw}
  \right) \frac{2}{\eps} \;,\no
\left(\delta Z_{\Zhat\Zhat} \right)^{\dive} &=&
  - \api \left(\frac{32 \sw^4 - 12 (2 \sw^2 - 1)}{9 \cw^2
  \sw^2} n - \frac{42 \cw^4 + 2 \cw^2 - 1}{6 \cw^2\sw^2}
  \right) \frac{2}{\eps} \;,\no
\left(\delta Z_{\What} \right)^{\dive} &=&
  - \api \left(\frac{4}{3\sw^2}n - \frac{43}{6\sw^2}
  \right) \frac{2}{\eps} \;,\no
\left(\delta Z_{\Hhat} \right)^{\dive} &=&
\left(\delta Z_{\chihat} \right)^{\dive} =
\left(\delta Z_{\phihat} \right)^{\dive} \no
  &=&
  - \api \left(\sum_{f}\frac{\Mf^2}{2 \MW^2 \sw^2} - \frac{2
  \cw^2 + 1}{4 \cw^2\sw^2} (\xiQ + 3)
  \right) \frac{2}{\eps} \; .
\eeqar
The divergent parts of the gauge-boson field renormalization constants
are independent of $\xiQ$ in accordance with the general
discussion given above.

The renormalized \ses\ are obtained by adding the counterterms
specified in \refeq{fr:AAct} -- \refeq{fr:CFFct} to the unrenormalized
\ses. It is evident from \refeqs{eq:ses} -- \refeqs{eq:fieldrc}
that although the field renormalization
constants cannot be chosen freely in the BFM, all renormalized
\ses\ are nevertheless finite.
Whereas in the conventional formalism the field renormalization
constants are adjusted in order to obtain finite \ses, this
happens automatically in the BFM as a consequence of the \WI.
The finiteness of the longitudinal parts of the gauge-boson
\ses\ and of the gauge-boson--scalar mixing energies follows
directly from the finiteness of the renormalized tadpole and
scalar \ses\ and the \WI\ \refeq{eq:segachi} -- \refeq{eq:seW2}.

A renormalization based on the on-shell definition of all
parameters can therefore consistently be used in the BFM. It
renders all vertex functions finite while respecting the full
gauge symmetry of the BFM.

Since the divergent parts of the unrenormalized \ses\ fulfill
the \WI\ by themselves, it is obvious that renormalization in
the minimal-subtraction scheme also preserves the symmetry of
the BFM.

As mentioned above, the on-shell renormalization in the BFM
fixes the residue of the photon propagator to unity. The propagators
of the other gauge bosons, scalars and
left-handed up-type fermions acquire residues
different from unity. This is similar to the minimal on-shell
scheme of the conventional formalism
and has to be corrected in the S-matrix
elements by a UV-finite wave-function renormalization.

The renormalization constants introduced in \refeqs{eq:renconsts1}
and \refeqs{eq:renconsts2} correspond to the
physical fields, i.e.~the mass and electric charge eigenstates $\Ahat,
\Zhat, \What^{\pm}$. Alternatively, one can introduce
renormalization constants in the symmetric
formulation (see e.g.~\citere{BHS}) resulting in the minimal
on-shell scheme.
In the bosonic sector these renormalization constants are given by
\beqar
\label{eq:rcBHS}
\What^a_0 &=& (Z_2^{\What})^{1/2} \What^a , \no
\Bhat_0   &=& (Z_2^{\Bhat})^{1/2} \Bhat , \no
\hat\Phi_0   &=& (Z^{\hat\Phi})^{1/2} \hat\Phi , \no
g_{2, 0} &=& Z_1^{\What} (Z_2^{\What})^{- 3/2} g_{2}
	     = Z_{g_2} g_{2} , \no
g_{1, 0} &=& Z_1^{\Bhat} (Z_2^{\Bhat})^{- 3/2} g_{1}
	     = Z_{g_1} g_{1} , \no
v_{ 0} &=& (Z^{\hat\Phi})^{1/2} (v - \de v), \no
\mu_{ 0}^2 &=& (Z^{\hat\Phi})^{-1} (\mu^2 - \de \mu^2), \no
\la_{ 0} &=& Z^\la(Z^{\hat\Phi})^{-2} \la .
\eeqar
In this formulation, the gauge symmetry of the BFM implies
in addition to $Z_1^{\Bhat} = Z_2^{\Bhat}$
\beqar \label{eq:rcsymm}
Z_1^{\What} &=&  Z_2^{\What} , \no
\de v &=& 0.
\eeqar
Thus, for both the isotriplet fields of SU$(2)_{\PW}$ and the
isosinglet field of U$(1)_{\rY}$ a QED-like relation
between coupling constant and field renormalization holds,  and
there is no renormalization of the vacuum expectation value other than
the one owing to the Higgs-field renormalization.
The other restrictions following from
\refeq{eq:delZZA} -- \refeqs{eq:zferm}
are already taken into account in the ansatz~\refeqs{eq:rcBHS}
for the field renormalization.
It is clear that also in this on-shell scheme the field renormalizations
of the gauge bosons are gauge-independent.
With the restrictions imposed
by the BFM, the two renormalization schemes become
in fact equivalent, \ie both schemes yield identical renormalized
Green functions.

We have derived the relations between the renormalization constants
from the back\-ground-field Ward identities given in
the last section.
As the gauge invariance of the effective action is directly
related to the gauge invariance of the classical
Lagrangian~\cite{Ab81}, those relations can also be inferred
directly from the Lagrangian. One can check that the relations
listed above are precisely those required to render the
renormalized classical Lagrangian ${\cal L}_{\mathrm{C}}$
gauge-invariant.

As a consequence of the relations \refeq{eq:delZZA} --
\refeq{eq:zferm},
the counterterm vertices of the background fields
have a much simpler structure than the ones in the conventional
formalism (see e.g.~\citere{Dehab}).
Their explicit form is given in the appendix.
Moreover,
all vertices resulting from an irreducible gauge-invariant part
of the Lagrangian and in particular all realizations of a
generic vertex,
e.g.~$\Vhat\Vhat\Vhat\Vhat$, are renormalized in the same way.

In the appendix we have listed the counterterms for all vertices
involving only background fields. These are
sufficient for the renormalization of all one-loop processes.
Through the parameter renormalizations and the renormalizations of the
background fields also the vertices containing both quantum and
background fields and the pure quantum-field vertices acquire
counterterms. These become relevant in higher orders. Their
explicit form can easily be obtained using~\refeqs{eq:renconsts1},
\refeqs{eq:renconsts2}
and the Feynman rules given in the appendix.

\section{PROPERTIES OF BFM VERTEX FUNCTIONS}

As mentioned above, the BFM vertex functions possess improved
theoretical properties compared to their conventional counterparts.
In previous treatments,
such properties were either explicitly enforced by construction
\cite{Ke89,Ku91} or could only be verified for specific examples
\cite{pinch,sirlin}. Since the properties could
not be derived from the theory, their theoretical understanding
remained unclear. Moreover, the
new ``vertex functions'' were obtained
by rearranging contributions between
different conventional Green functions.
The field-theoretical meaning of these objects is obscure.
In the BFM, the background-field vertex functions themselves exhibit
the improved properties.
As will be illustrated in this section, these properties
can be directly deduced from
the \WI\ discussed in \refse{sec:WI}. The Ward identities
are a direct consequence of the background-field gauge invariance
and are valid  independent of the value of the quantum gauge parameter
$\xi_Q$.
Consequently, the properties of the BFM vertex functions
following from these identities also
hold for arbitrary $\xi_Q$.

We first consider the
fermion--gauge-boson vertex functions. In~\citere{sirlin} it was
found by explicit calculation that in the pinch technique
the one-loop fermion--gauge-boson vertex
functions are UV-finite when the fermion field
renormalization has been added. In the BFM, this fact is an
obvious consequence of the relations between the renormalization
constants derived in the last section. As follows from
\refeq{fr:Vff}, the counterterm for the
$\Vhat\bar{F}F$-vertex is solely given by the fermion
field renormalization. Adding it to the vertex function
evidently cancels the UV divergence. Obviously, this fact holds
for all values of the quantum gauge parameter $\xi_Q$.
{}From the counterterm structure given in the appendix
similar conclusions can be drawn for other vertex functions.
In particular, the $\Vhat\What\What$ and $\Vhat\Vhat'\What\What$
vertices become UV-finite after adding the field renormalization
of two $\What$ fields as can be read from (\ref{fr:vvvv}) and
(\ref{fr:vvv}). In~\citere{sirlin} it was also noted that the one-loop
fermion--photon vertex functions including fermion field
renormalization vanish at zero momentum transfer of
the photon. In the BFM, the inclusion of the fermion field
renormalization amounts to the complete renormalization of this vertex.
But the renormalized vertex correction vanishes owing to the
renormalization condition
for the electric charge.

Next, we investigate the asymptotic behavior of the gauge-boson
\ses\ in the BFM for $|q^2| \to \infty$. In~\citere{bgfproc}, the
explicit one-loop  result for the leading logarithms
of the bosonic contributions
to the gauge-boson \ses\ in the BFM
has been given
showing that their coefficients are independent of
$\xi_Q$. However, this feature can also be deduced from the Ward
identities as follows. In dimensional
regularization the unrenormalized one-loop \ses\ obey
$(\Vhat,\Vhat'=\Ahat,\Zhat,\What)$
\beq
\Pi^{\Vhat\Vhat'}(q^2) =
\frac{\Sigma^{\Vhat\Vhat'}_\rT(q^2)-\Sigma^{\Vhat\Vhat'}_\rT(0)}{q^2} =
g_{\Vhat\Vhat'}^2 \,
\mu^{\eps} \left(- c_{\Vhat\Vhat'} \frac{2}{\eps} + \mbox{UV-finite
terms} \right), % + \O\left(\frac{1}{q^2}\right),
\eeq
where
$c_{\Vhat\Vhat'}$ is a $q^2$-independent coefficient, which can
be read off from \refeqs{eq:ses},
$g_{\Ahat\Ahat}=e$, $g_{\What\What}=g_2=e/\sw$,
$g_{\Zhat\Zhat}=e/(\cw\sw)$,
$g_{\Ahat\Zhat}=e/\sqrt{\cw\sw}$,
and $\mu$
is a mass parameter necessary to keep $g_{\Vhat\Vhat'}^2$ dimensionless.
In the limit $|q^2| \to \infty$ all masses can be neglected and on
dimensional grounds the \ses\ behave as
\beq
\Pi^{\Vhat\Vhat'}(q^2) \;\mbox{$\asymp{|q^2|\to\infty}$}\;
g_{\Vhat\Vhat'}^2 \left(- c_{\Vhat\Vhat'} \frac{2}{\eps}
+ c_{\Vhat\Vhat'} \log\frac{|q^2|}{\mu^2} + \mbox{UV-finite
constant} \right) .
\eeq
Using the identities
\beq \label{eq:ZVVdef}
\left(\delta Z_{\Vhat\Vhat}\right)^{\rm div} =
- \left(\Pi^{\Vhat\Vhat}(q^2) \right)^{\rm div} , \quad
\left(\delta Z_{\Ahat\Zhat}\right)^{\rm div} =
- 2 \left(\Pi^{\Ahat\Zhat}(q^2) \right)^{\rm div}
\label{eq:divdelaa}
\eeq
we can identify the divergent parts of $\delta Z_{\Vhat\Vhat}$
and $\delta Z_{\Ahat\Zhat}$ as
\beq \label{eq:ZVVres}
\left(\delta Z_{\Vhat\Vhat}\right)^{\rm div} =
g_{\Vhat\Vhat}^2 c_{\Vhat\Vhat} \frac{2}{\eps} , \quad
\left(\delta Z_{\Ahat\Zhat}\right)^{\rm div} =
2 g_{\Ahat\Zhat}^2 c_{\Ahat\Zhat} \frac{2}{\eps} .
\eeq
We found in \refse{se:renco} that the field renormalization
constants for the gauge bosons and thus
$\left(\delta Z_{\Vhat\Vhat'}\right)^{\rm div}$
are gauge-independent. As a consequence,
also the coefficients $c_{\Vhat\Vhat'}$ of the leading logarithms of
$\Sigma^{\Vhat\Vhat'}_\rT(q^2)$ are independent of $\xi_Q$.

In \citere{Ab81} it has been shown for QCD that in the BFM
the $\beta$-function of the gauge coupling is related
to the anomalous dimension and thus to the field renormalization
constant of the gauge boson.
The same applies to the SM as well.
The relation $Z_e = Z_{\Ahat\Ahat}^{-1/2}$ implies for the
$\be$-function associated with the electromagnetic coupling
in the minimal-subtraction scheme
\beq
\be_e(e) = c_{\Ahat\Ahat} e^3 + \O(e^5) ,
\eeq
i.e.~in analogy to QED, the coefficient of the
leading logarithm of the photon \se\ in the BFM equals
the coefficient of the one-loop $\be$-function.
Analogously, the relation
$Z_{g_2} = (Z_2^{\What})^{-1/2} = Z_{\What}^{-1/2}$,
which can be inferred from \refeqs{eq:delZB} and
\refeqs{eq:rcsymm},
yields for the charged-current coupling
\beq
\be_{g_2}(g_2) = c_{\What\What} g_2^3 + \O(g_2^5).
\eeq

The fact that the coefficients of the leading logarithms of the \ses\
equal the coefficients of the $\be$-functions implies
that the asymptotic behavior of effective
coupling constants $e^2(q^2)$ and $g_2^2(q^2)$
defined via Dyson summation of \ses\
(see e.g. \citeres{Ke89,Ku91,sirlin}) is governed by the
renormalization
group. As a consequence, we can introduce running couplings as follows
\beqar
\label{eq:runcoupl}
e^2(q^2) &=& \frac{e_0^2}{1 + \Re \Pi^{\Ahat\Ahat}(q^2)}
	 = \frac{e^2}{1 + \Re \Pi^{\Ahat\Ahat,\ren}(q^2)} , \no
g_2^2(q^2) &=& \frac{g_{2,0}^2}{1 + \Re\Pi^{\What\What}(q^2)}
         = \frac{g_2^2}{1 + \Re\Pi^{\What\What,\ren}(q^2)} ,
\eeqar
where the quantities on the right-hand side are the renormalized
ones and the second equality holds because of
$Z_e = Z_{\Ahat\Ahat}^{-1/2}$ and $Z_{g_2} = Z_{\What}^{-1/2}$,
respectively.
As these running couplings can be expressed in terms of bare
quantities, %note that the effective coupling constant defined in this way
they are manifestly renormalization-scheme independent in the BFM .
Asymptotically these couplings are equivalent to the ones defined in
\citeres{Ke89,Ku91,sirlin}, but for finite values of $q^2$ there are
differences.%
\footnote{Those differences also exist between the different
formulations of the previous treatments.}
Moreover, the running couplings \refeq{eq:runcoupl} depend
on $\xi_Q$ in the non-asymptotic region. This indicates that
any definition of running couplings via Dyson summation of \ses\
that take into account mass effects is not unique but a matter
of convention.
This arbitrariness is made transparent in the BFM and has to be
taken into account when considering applications.

We can define a running $\sw(q^2)$ as the ratio of the
electromagnetic and charged-current running coupling constants
\beq
\sw ^2(q^2) = \frac{e^2(q^2)}{g_2^2(q^2)}
= \frac{e^2}{g_2^2}\frac{1 + \Re\Pi^{\What\What,\ren}(q^2)}
                        {1 + \Re \Pi^{\Ahat\Ahat,\ren}(q^2)}.
\eeq
In the leading-logarithmic approximation this can be written as
\beq
\sw ^2(q^2) \;\mbox{$\asymp{|q^2|\to\infty}$}\;
\sw ^2 \left(1 - \frac{\cw}{\sw}
\Re\Pi^{\Ahat\Zhat,\ren}(q^2)\right)
+ \O(e^4).
\eeq
This resembles the definition of a running $\sw^2(q^2)$ used for
example in~\citere{Hollik}.

\section{CONCLUSION}

In this paper we have studied the application of the BFM to the
electroweak SM. We have given the full Lagrangian for the SM
and indicated how the gauge-invariant effective action of the
BFM and the S matrix are constructed. A complete set of Feynman
rules for arbitrary values of a quantum gauge parameter has been
listed including all counterterms necessary for one-loop
calculations.

We have shown that the gauge invariance of the BFM implies
simple QED-like \WI. They have been discussed in
comparison with the Slavnov--Taylor identities of the conventional
formalism. As a consequence of the \WI, the vertex functions
in the BFM possess improved
theoretical properties compared to their conventional counterparts.
In particular, this has been worked out for the example
of running couplings directly defined via Dyson summation. In
contrast to the conventional formalism, their asymptotic behavior
is automatically governed by
the renormalization group and independent of the quantum gauge
parameter. In comparison to former treatments like the pinch
technique,
where desirable properties of Green functions could
only be verified by explicit computation, the BFM offers a
well-suited framework for studying the properties of off-shell
Green functions by relating them to the gauge invariance of the
effective action.

Moreover, practical calculations of S-matrix elements simplify
considerably in the BFM. The freedom to choose an appropriate
gauge, e.g.~the unitary gauge, for the background fields
independently of the quantum gauge fixing allows to reduce the
number of contributing Feynman diagrams drastically.
In addition, also the evaluation of loop diagrams simplifies. This holds
in particular in the 't Hooft--Feynman gauge for the quantum fields.

When considering applications of the BFM in the SM it is
particularly important to establish %a procedure for
a consistent
renormalization which does not violate the explicit gauge
invariance, i.e.~which does not alter the form of the \WI.
This has been done starting from two different
renormalization schemes, a complete %on-shell scheme
and a minimal on-shell scheme. We have shown that
the gauge symmetry imposes relations between field
renormalization constants and the renormalization constants
of the SM parameters,
i.e.~electric charge and particle masses.
It was
pointed out that even with this reduced set of independent
renormalization constants all Green functions of the SM become
finite. This has been verified explicitly at one-loop order by
calculating the relevant quantities.
The renormalization constants of the physical parameters are
still independent of each other
so that all on-shell parameter renormalization
conditions can be maintained. Thus, the on-shell scheme is
compatible with the symmetries of the BFM.
Furthermore, it is obvious that the
same holds for the
minimal-subtraction scheme.

As a consequence of gauge invariance, the renormalization in
the BFM drastically simplifies compared to the conventional
formalism both technically and conceptually. In the BFM, much
less independent renormalization constants are needed and the
counterterms have a much simpler structure. All realizations of
a generic vertex have one single universal counterterm. If
charge and particle masses are identified with their physical
values, the field renormalizations of all gauge bosons
become gauge-independent.

\section*{Acknowledgement}
We thank M.~B\"ohm and H.~Spiesberger for useful discussions.

\section*{Note added}
Shortly before completion of this paper we became
aware of a simultaneous work~\cite{Li} focussing on the
renormalization of the electroweak SM (omitting fermions)
in the BFM. As
in this reference the residue of the Higgs field is required
to be unity, in contrast to our result~\refeq{eq:rcsymm}
a nonzero (but nevertheless finite)
correction $\delta v$ to the vacuum expectation value of the
Higgs field is needed.
This violates the na\"\i{}ve Ward identities and is cured
in~\citere{Li} by including $\delta v$ into the renormalized Ward
identities. Since the renormalization in the BFM necessarily
involves fields whose residues differ from unity we find it
preferable to carry it out in such a way that the explicit gauge
invariance and correspondingly the form of the Ward identities
is retained.
Furthermore, we disagree with
the conclusion of~\citere{Li} that the Landau
gauge would be enforced for the background fields.
In fact, we do not find any reason that would require this restriction.

\appendix
\section{FEYNMAN RULES IN THE BACKGROUND-FIELD METHOD}
\label{app:Frules}

\unitlength 1pt

\savebox{\Gr}(48,0)[bl]
{ \put(23,0){\vector(1,0){3}} \multiput(0,0)(4.8,0){11}{\circle*1{1}} }
\savebox{\Gtbr}(32,48)[bl]
{ \multiput(0,24)(4,3){9}{\circle*{1}}
 \multiput(0,24)(4,-3){9}{\circle*{1}}
 \put(16,12){\vector(-4,3){3}} \put(16,36){\vector(4,3){3}} }

\savebox{\wigr}(12,0)[bl]
   {\bezier{20}(0,0)(3, 4)(6,0)
    \bezier{20}(6,0)(9,-4)(12,0)}
\savebox{\Vr}(48,0)[bl]{\multiput(0,0)(12,0){4}{\usebox{\wigr}}}
\savebox{\wigur}(8,6)[bl]
   {\bezier{20}(0,0)(1,4)(4,3)
    \bezier{20}(4,3)(7,2)(8,6)}
\savebox{\Vtr}(32,24)[bl]{\multiput(0,0)(8,6){4}{\usebox{\wigur}}}
\savebox{\wigdr}(8,6)[bl]
   {\bezier{20}(0,0)(1,-4)(4,-3)
    \bezier{20}(4,-3)(7,-2)(8,-6)}
\savebox{\Vbr}(32,24)[bl]{\multiput(0,18)(8,-6){4}{\usebox{\wigdr}}}
\savebox{\Vtbr}(32,48)[bl]{\put(00,24){\usebox{\Vtr}}
                           \put(00,00){\usebox{\Vbr}}}

\savebox{\Sr}(48,0)[bl]
{ \multiput(0,0)(12.666,0){4}{\line(4,0){10}} }
\savebox{\Str}(32,24)[bl]
{ \multiput(0,0)(11.5,8.625){3}{\bezier{30}(0,0)(4,3)(9,6.75) } }
\savebox{\Sbr}(32,24)[bl]
{ \multiput(0,24)(11.5,-8.625){3}{\bezier{30}(0,0)(4,-3)(9,-6.75) } }
\savebox{\Stbr}(32,48)[bl]
{\put(00,24){\usebox{\Str}}
\put(00,00){\usebox{\Sbr}}}

\savebox{\Fr}(48,0)[bl]
{ \put(0,0){\vector(1,0){26}} \put(24,0){\line(1,0){24}} }
\savebox{\Ftr}(32,24)[bl]
{ \put(0,0){\vector(4,3){18}} \put(16,12){\line(4,3){16}} }
\savebox{\Fbr}(32,24)[bl]
{ \put(32,0){\vector(-4,3){19}} \put(16,12){\line(-4,3){16}} }
\savebox{\Ftbr}(32,48)[bl]
{\put(00,24){\usebox{\Ftr}}
\put(00,00){\usebox{\Fbr}}}

\renewcommand{\arraystretch}{1.4}

In this appendix we list the Feynman rules of the SM in the
BFM for an arbitrary quantum gauge parameter
$\xiQ = \xiQ^W = \xiQ^B$. We write down generic Feynman rules
for all vertices and give the possible actual insertions. We use
here the shorthand notation
\beq
c = \cw,\qquad s= \sw.
\eeq

{}From the Feynman rules given here, the vertex
functions corresponding to the gauge-invariant effective action
of the BFM can be calculated. No gauge-fixing
term is included for the background fields.
Such a term is only relevant for the construction of connected Green
functions and S-matrix elements from the vertex functions. It
can be chosen independently from the gauge-fixing of the quantum
fields. If a linear gauge is used, only the
propagators of the background fields are affected.
In a background $R_\xi$ gauge,
the background-field  propagators take the same form
as the quantum-field propagators given below with $\xi_Q$ replaced by the
background gauge parameter $\xi_B$. Note, however, that it is
preferable to use a more convenient gauge for the background fields
like the unitary gauge.

We first list the vertices containing only background fields
including counterterms.
In lowest order, these vertices are identical to the ones in the
conventional formalism
(see e.g.~\citere{Dehab}). Their
counterterms, however, have a much simpler structure. Note that
in the BFM apart from the two-point functions each generic vertex
has a universal counterterm.
As mentioned above, these
counterterms are sufficient for the renormalization of all
one-loop processes.

In the vertices all momenta and fields are considered as incoming.

% FLEQN DOCUMENT-STYLE OPTION - released 04 November 1991
%    for LaTeX version 2.09
% Copyright (C) 1989,1991 by Leslie Lamport

\typeout{Document style option `fleqn' - Released 04 Nov 91}

\def\[{\relax\ifmmode\@badmath\else
 \begin{trivlist}%
 \@beginparpenalty\predisplaypenalty
 \@endparpenalty\postdisplaypenalty
 \item[]\leavevmode
 \hbox to\linewidth\bgroup $\m@th\displaystyle
 \hskip\mathindent\bgroup\fi}

\def\]{\relax\ifmmode \egroup $\hfil
       \egroup \end{trivlist}\else \@badmath \fi}

\def\equation{\@beginparpenalty\predisplaypenalty
  \@endparpenalty\postdisplaypenalty
\refstepcounter{equation}\trivlist \item[]\leavevmode
  \hbox to\linewidth\bgroup $\m@th% $ TO MAKE DOLLAR NESTING OK
  \displaystyle
\hskip\mathindent}

\def\endequation{$\hfil
           \displaywidth\linewidth\@eqnnum\egroup \endtrivlist}

\def\eqnarray{\stepcounter{equation}\let\@currentlabel=\theequation
\global\@eqnswtrue
\global\@eqcnt\z@\tabskip\mathindent\let\\=\@eqncr
\abovedisplayskip\topsep\ifvmode\advance\abovedisplayskip\partopsep\fi
\belowdisplayskip\abovedisplayskip
\belowdisplayshortskip\abovedisplayskip
\abovedisplayshortskip\abovedisplayskip
$$\m@th\halign
to\linewidth\bgroup\@eqnsel\hskip\@centering$\displaystyle\tabskip\z@
  {##}$&\global\@eqcnt\@ne \hskip 2\arraycolsep \hfil${##}$\hfil
  &\global\@eqcnt\tw@ \hskip 2\arraycolsep $\displaystyle{##}$\hfil
   \tabskip\@centering&\llap{##}\tabskip\z@\cr}

\def\endeqnarray{\@@eqncr\egroup
      \global\advance\c@equation\m@ne$$\global\@ignoretrue
      }

\newdimen\mathindent
\mathindent = \leftmargini

%\endinput

\addtolength{\mathindent}{-4pt}

\begin{itemize}
\item{tadpole:
\beq
\barr{l}
\framebox {
\begin{picture}(72,27)
\put(50,15){\makebox(10,10)[bl]{$\Hhat$}}
\put(7.5,6){\line(3,4){9}}
\put(7.5,18){\line(3,-4){9}}
\put(12,12){\usebox{\Sr}}
\end{picture} }
\earr
\barr{l}
\disp = i\delta t.%\frac{2s}{e}\MW\MH^{2}\frac{\delta t}{t}.
\earr
\eeq
}
\item{
{\samepage
\noindent
$\Vhat\Vhat$ counterterm:
\beq
\label{fr:AAct}
\barr{l}
\framebox{
\begin{picture}(120,35)
\put(100,19){\makebox(10,20)[bl]{$\Vhat_{2,\nu}$}}
\put(0,19){\makebox(10,20)[bl] {$\Vhat_{1,\mu},k$}}
\put(55.5,6){\line(3,4){9}}
\put(55.5,18){\line(3,-4){9}}
\put(12,10){\usebox{\Vr}}
\put(60,10){\usebox{\Vr}}
\end{picture} }
\earr
\barr{l}
= i \Bigl[ ( - g_{\mu\nu} k^2 + k_{\mu} k_{\nu} ) C_{1}  +
g_{\mu\nu} C_{2} \Bigr]
\earr
\eeq
}
with the actual values of $\Vhat_1$, $\Vhat_2$ and $C_{1}$, $C_{2}$
\beq
\begin{array}{|r||c|c|c|c|} \hline
\Vhat_1\Vhat_2 & \What^+\What^- & \Zhat\Zhat & \Ahat\Zhat
& \Ahat\Ahat\\ \hline
C_1 & \delta Z_{\What} & \delta Z_{\Zhat\Zhat} & \frac{1}{2}
 \delta Z_{\Ahat\Zhat} & \delta Z_{\Ahat\Ahat} \\ \hline
C_2 & \MW^2 \delta Z_{\What} + \delta \MW^2 &
 \MZ^2 \delta Z_{\Zhat\Zhat} + \delta \MZ^2 & 0 & 0 \\ \hline
\end{array}
\eeq
}

{\item
{\samepage
\noindent
$\Vhat{\hat S}$ counterterm:
\beq
\barr{l}
\framebox{
\begin{picture}(120,35)
\put(100,19){\makebox(10,20)[bl]{${\hat S}$}}
\put(0,19){\makebox(10,20)[bl] {$\Vhat_{\mu},k$}}
\put(55.5,6){\line(3,4){9}}
\put(55.5,18){\line(3,-4){9}}
\put(12,10){\usebox{\Vr}}
\put(60,12){\usebox{\Sr}}
\end{picture} }
\earr
\barr{l}
= i k_{\mu} C \delta Z_{\Hhat}
\earr
\eeq
}
with the actual values of $\Vhat$, $\Shat$ and $C$
\beq
\barr{|r||c|c|c|} \hline
\Vhat{\hat S} & \What^\pm\phihat^\mp & \Zhat\chihat\\ \hline
C & \pm\MW & i \MZ\\ \hline
\earr
\eeq
}

{\item
{\samepage
\noindent
${\hat S} {\hat S}$ counterterm:
\beq
\barr{l}
\framebox{
\begin{picture}(120,30)
\put(100,18){\makebox(10,20)[bl]{${\hat S}_{2}$}}
\put(0,18){\makebox(10,20)[bl] {${\hat S}_{1},k$}}
\put(55.5,6){\line(3,4){9}}
\put(55.5,18){\line(3,-4){9}}
\put(12,12){\usebox{\Sr}}
\put(60,12){\usebox{\Sr}}
\end{picture} }
\earr
\barr{l}
= i\Bigl[\delta Z_{\Hhat}k^{2} - C \Bigr]
\earr
\eeq
}
with the actual values of $\Shat_{1}$, $\Shat_{2}$ and
$C_{1}$, $C_{2}$
\beq
\barr{|r||c|c|} \hline
{\hat S}_1{\hat S}_2 & \Hhat\Hhat & \chihat \chihat, \phihat
 \phihat\\ \hline
C & \MH^2 \delta Z_{\Hhat} + \delta \MH^2 &
 - \frac{e}{2 s} \frac{\delta t}{\MW} \\ \hline
\earr
\eeq
}

{\item
{\samepage
\noindent
$F\bar{F}$ counterterm:
\beq
\barr{l}
\framebox{
\begin{picture}(120,30)
\put(100,18){\makebox(10,20)[bl]{$\bar{F}_{2}$}}
\put(0,18){\makebox(10,20)[bl] {$F_{1},p$}}
\put(55.5,6){\line(3,4){9}}
\put(55.5,18){\line(3,-4){9}}
\put(12,12){\usebox{\Fr}}
\put(60,12){\usebox{\Fr}}
\end{picture} }
\earr
\barr{l}
= i\Bigl[C_{\rL} \ps \omega_{-} +
C_{\rR} \ps \omega_{+} - C_{\rS} \Bigr]
\earr
\eeq
}
with the actual values of $F_{1}$, $\bar{F}_{2}$ and
$C_{\rL}$, $C_{\rR}$, $C_{\rS}$
\beq
\label{fr:CFFct}
\barr{|r||c|} \hline
F_{1} \bar{F}_{2} & f \bar{f}\\ \hline
C_{\rL} & \delta Z_f^{\rL}\\ \hline
C_{\rR} & \delta Z_f^{\rR}\\ \hline
C_{\rS} & m_f \frac{1}{2} \left(\delta Z_f^{\rL} +
\delta Z_f^{\rR}\right) + \delta m_f\\ \hline
\earr
\eeq
}

{\item
{\samepage
$\Vhat\Vhat\Vhat\Vhat$ coupling:
\beq
\label{fr:vvvv}
\barr{l}
\framebox{
\begin{picture}(96,82)(0,-4)
\put(6,65){\makebox(10,20)[bl]{$\Vhat_{1,\mu}$}}
\put(72,65){\makebox(10,20)[bl]{$\Vhat_{3,\rho}$}}
\put(6,-4){\makebox(10,20)[bl]{$\Vhat_{2,\nu}$}}
\put(72,-4){\makebox(10,20)[bl]{$\Vhat_{4,\sigma}$} }
\put(48,36){\circle*{4}}
\put(16,12){\usebox{\Vtr}}
\put(16,36){\usebox{\Vbr}}
\put(48,12){\usebox{\Vtbr}}
\end{picture} }
\earr
\barr{l}
= ie^{2}C \Bigl[2g_{\mu\nu}g_{\sigma \rho } -
g_{\nu\rho}g_{\mu \sigma } - g_{\rho\mu}g_{\nu \sigma }\Bigr]
(1 + \delta Z_{\What})
\earr
\eeq
}
with the actual values of $\Vhat_1$, $\Vhat_2$, $\Vhat_3$, $\Vhat_4$
and $C$
\beq
\label{fr:Cvvvv}
\barr{|r||c|c|c|c|} \hline
\Vhat_1\Vhat_2\Vhat_3\Vhat_4 & \What^+\What^+\What^-\What^- &
\What^+\What^-\Zhat\Zhat & \What^+\What^-\Ahat\Zhat &
\What^+\What^-\Ahat\Ahat\\ \hline
C & \frac{1}{s^2} & -\frac{c^2}{s^2} & \frac{c}{s} & -1\\ \hline
\earr
\eeq
}

{\item
{\samepage
$\Vhat\Vhat\Vhat$ coupling:
\beq
\label{fr:vvv}
\barr{l}
\framebox{
\begin{picture}(96,82)(0,-4)
\put(60,65){\makebox(10,20)[bl]{$\Vhat_{2,\nu},k_{2}$}}
\put(0,43){\makebox(10,20)[bl] {$\Vhat_{1,\mu},k_{1}$}}
\put(60,-4){\makebox(10,20)[bl] {$\Vhat_{3,\rho},k_{3}$}}
\put(48,36){\circle*{4}}
\put(0,34){\usebox{\Vr}}
\put(48,12){\usebox{\Vtbr}}
\end{picture} }
\earr
\barr{l}
= -ieC \Bigl[g_{\mu \nu }(k_{2}-k_{1})_{\rho}
+g_{\nu\rho}(k_{3}-k_{2})_{\mu}
\\ \hphantom{= -ieC \Bigl[}
+g_{\rho\mu}(k_{1}-k_{3})_{\nu}\Bigr]
(1 + \delta Z_{\What})
\earr
\eeq
}
with the actual values of $\Vhat_1$, $\Vhat_2$, $\Vhat_3$ and $C$
\beq
\label{fr:Cvvv}
\barr{|r||c|c|} \hline
\Vhat_1\Vhat_2\Vhat_3 & \Ahat\What^+\What^- & \Zhat\What^+\What^-\\
\hline
C & 1 & - \frac{c}{s}\\ \hline
\earr
\eeq
}

{\item
{\samepage
$\Shat\Shat\Shat\Shat$ coupling:
\beq
\label{fr:ssss}
\barr{l}
\framebox{
\begin{picture}(96,82)(0,-4)
\put(9,65){\makebox(10,20)[bl]{$\Shat_{1}$}}
\put(75,65){\makebox(10,20)[bl]{$\Shat_{3}$}}
\put(9,-4){\makebox(10,20)[bl]{$\Shat_{2}$}}
\put(75,-4){\makebox(10,20)[bl]{$\Shat_{4}$}}
\put(48,36){\circle*{4}}
\put(16,12){\usebox{\Str}}
\put(16,36){\usebox{\Sbr}}
\put(48,12){\usebox{\Stbr}}
\end{picture} }
\earr
\barr{l}  \disp
= ie^{2}C \Bigl[1 + \frac{\delta \MH^2}{\MH^2} + \frac{e}{2 s}
		\frac{\delta t}{\MW \MH^2} + \delta Z_{\Hhat}
          \Bigr]
\earr
\eeq
}
with the actual values of $\Shat_1$, $\Shat_2$, $\Shat_3$, $\Shat_4$ and $C$
\beq
\label{fr:Cssss}
\barr{|r||c|c|c|} \hline
\Shat_1\Shat_2\Shat_3\Shat_4 & \Hhat\Hhat\Hhat\Hhat,
\chihat\chihat\chihat\chihat &
\Hhat\Hhat\chihat\chihat, \Hhat\Hhat\phihat^+\phihat^-,
\chihat\chihat\phihat^+\phihat^- &
\phihat^+\phihat^-\phihat^+\phihat^- \\ \hline
C & -\frac{3}{4 s^2} \frac{\MH^2}{\MW^2} &
-\frac{1}{4 s^2} \frac{\MH^2}{\MW^2} &
-\frac{1}{2 s^2} \frac{\MH^2}{\MW^2}\\ \hline
\earr
\eeq
}

{\item
{\samepage
$\Shat\Shat\Shat$ coupling:
\beq
\barr{l}
\framebox{
\begin{picture}(96,82)(0,-4)
\put(75,65){\makebox(10,20)[bl]{$\Shat_{2}$}}
\put(0,42){\makebox(10,20)[bl]{$\Shat_{1}$}}
\put(75,-3){\makebox(10,20)[bl]{$\Shat_{3}$}}
\put(48,36){\circle*{4}}
\put(0,36){\usebox{\Sr}}
\put(48,12){\usebox{\Stbr}}
\end{picture} }
\earr
\barr{l} \disp
= ieC \Bigl[1 + \frac{\delta \MH^2}{\MH^2} + \frac{e}{2 s}
    	    \frac{\delta t}{\MW \MH^2} + \delta Z_{\Hhat}
      \Bigr]
\earr
\eeq
}
with the actual values of $\Shat_1$, $\Shat_2$, $\Shat_3$ and $C$
\beq
\barr{|r||c|c|} \hline
\Shat_1\Shat_2\Shat_3 & \Hhat\Hhat\Hhat &
\Hhat\chihat\chihat, \Hhat\phihat^+\phihat^- \\ \hline
C & -\frac{3}{2 s} \frac{\MH^2}{\MW} &
-\frac{1}{2 s} \frac{\MH^2}{\MW}\\ \hline
\earr
\eeq
}

{\item
{\samepage
$\Vhat\Vhat\Shat\Shat$ coupling:
\beq
\barr{l}
\framebox{
\begin{picture}(96,82)(0,-4)
\put(6,65){\makebox(10,20)[bl]{$\Vhat_{1,\mu}$}}
\put(6,-4){\makebox(10,20)[bl]{$\Vhat_{2,\nu}$}}
\put(75,65){\makebox(10,20)[bl]{$\Shat_{1}$}}
\put(75,-4){\makebox(10,20)[bl]{$\Shat_{2}$}}
\put(48,36){\circle*{4}}
\put(16,12){\usebox{\Vtr}}
\put(16,36){\usebox{\Vbr}}
\put(48,12){\usebox{\Stbr}}
\end{picture} }
\earr
\barr{l}
= ie^{2}g_{\mu\nu}C (1 + \delta Z_{\Hhat})
\earr
\eeq
}
with the actual values of $\Vhat_1$, $\Vhat_2$, $\Shat_1$, $\Shat_2$ and $C$
\bma
\barr{|r||c|c|c|c|c|} \hline
\Vhat_1\Vhat_2\Shat_1\Shat_2 & \Zhat\Zhat\Hhat\Hhat &
\What^+\What^-\Hhat\Hhat, \What^+\What^-\phihat^+\phihat^- &
\Ahat\Ahat\phihat^+\phihat^- & \Zhat\Ahat\phihat^+\phihat^- &
\Zhat\Zhat\phihat^+\phihat^-\\
& \Zhat\Zhat\chihat\chihat &
\What^+\What^-\chihat\chihat\hfill & & & \\ \hline
C  & \frac{1}{2 c^2 s^2} & \frac{1}{2 s^2} & 2 &
-\frac{c^2 - s^2}{c s} & \frac{(c^2 - s^2)^2}{2 c^2 s^2}\\ \hline
\earr
\ema
and
\beq
\barr{|r||c|c|c|c|c|} \hline
\Vhat_1\Vhat_2\Shat_1\Shat_2 &
\What^{\pm}\Ahat\phihat^{\mp}\Hhat &
\What^{\pm}\Ahat\phihat^{\mp}\chihat &
\What^{\pm}\Zhat\phihat^{\mp}\Hhat &
\What^{\pm}\Zhat\phihat^{\mp}\chihat \\ \hline
C & -\frac{1}{2 s} & \mp\frac{i}{2 s} &
-\frac{1}{2 c} & \mp \frac{i}{2c}\\ \hline
\earr
\eeq
}

{\item
{\samepage
$\Vhat\Shat\Shat$ coupling:
\beq
\label{fr:vss}
\barr{l}
\framebox{
\begin{picture}(96,82)(0,-4)
\put(67,65){\makebox(10,20)[bl]{$\Shat_{1},k_{1}$}}
\put(0,42){\makebox(10,20)[bl]{$\Vhat_{\mu}$}}
\put(67,-2){\makebox(10,20)[bl]{$\Shat_{2},k_{2}$}}
\put(48,36){\circle*{4}}
\put(0,34){\usebox{\Vr}}
\put(48,12){\usebox{\Stbr}}
\end{picture} }
\earr
\barr{l}
= ieC(k_{1}-k_{2})_{\mu} (1 + \delta Z_{\Hhat})
\earr
\eeq
}
with the actual values of $\Vhat$, $\Shat_1$, $\Shat_2$ and $C$
\beq \label{fr:Cvss}
\barr{|r||c|c|c|c|c|} \hline
\Vhat\Shat_1\Shat_2 & \Zhat\chihat\Hhat & \Ahat\phihat^+\phihat^- &
\Zhat\phihat^+\phihat^- & \What^{\pm}\phihat^{\mp}\Hhat &
\What^{\pm}\phihat^{\mp}\chihat\\ \hline
C & -\frac{i}{2 c s} & -1 & \frac{c^2 - s^2}{2 c s} &
\mp \frac{1}{2 s} & - \frac{i}{2 s}\\ \hline
\earr
\eeq
}

{\item
{\samepage
$\Shat\Vhat\Vhat$ coupling:
\beq
\barr{l}
\framebox{
\begin{picture}(96,82)(0,-4)
\put(72,65){\makebox(10,20)[bl]{$\Vhat_{1,\mu}$}}
\put(72,-4){\makebox(10,20)[bl] {$\Vhat_{2,\nu}$}}
\put(0,42){\makebox(10,20)[bl]{$\Shat$}}
\put(48,36){\circle*{4}}
\put(0,36){\usebox{\Sr}}
\put(48,12){\usebox{\Vtbr}}
\end{picture} }
\earr
\barr{l}
= ieg_{\mu\nu}C (1 + \delta Z_{\Hhat})
\earr
\eeq
}
with the actual values of $\Shat$, $\Vhat_1$, $\Vhat_2$ and $C$
\beq
\barr{|r||c|c|c|c|} \hline
\Shat\Vhat_1\Vhat_2 & \Hhat\Zhat\Zhat & \Hhat\What^+\What^- &
\phihat^{\pm}\What^{\mp}\Ahat & \phihat^{\pm}\What^{\mp}\Zhat  \\ \hline
C & \frac{1}{c^2s} \MW & \frac{1}{s} \MW & -\MW &
-\frac{s}{c} \MW\\ \hline
\earr
\eeq
}

{\item
{\samepage
$\Vhat \bar{F}F$ coupling:
\beq
\label{fr:Vff}
\barr{l}
\framebox{
\begin{picture}(96,82)(0,-4)
\put(75,65){\makebox(10,20)[bl]{$\bar{F}_{1}$}}
\put(0,42){\makebox(10,20)[bl]{$\Vhat_{\mu}$}}
\put(75,-2){\makebox(10,20)[bl]{$F_{2}$}}
\put(48,36){\circle*{4}}
\put(0,34){\usebox{\Vr}}
\put(48,12){\usebox{\Ftbr}}
\end{picture} }
\earr
\barr{l}
= ie\gamma_{\mu} \Bigl[ C_{\rL}\omega_{-} (1 + \delta Z^{\rL}_{F_1}) \\
{} \quad \phantom{\ga_\mu} + C_{\rR}\omega_{+}
\Bigr(1 + \frac{1}{2}(\de Z^{\rR}_{F_1} + \de Z^{\rR}_{F_2})\Bigl) \Bigr]
\earr
\eeq
}
with the actual values of $\Vhat$, $\bar{F}_{1}$, $F_{2}$ and
$C_{\rR}$, $C_{\rL}$
\beq \label{fr:CVff}
\barr{|r||c|c|c|} \hline
\Vhat\bar{F}_{1} F_{2} & \Ahat\bar{f} f & \Zhat\bar{f} f &
\What^+\bar{f}_u f_d , \What^-\bar{f}_d f_u\\ \hline
C_{\rL} & - Q_f & \frac{I_{\rw, f}^3 - s^2 Q_f}{c s} &
\frac{1}{\sqrt{2} s}\\ \hline
C_{\rR} & - Q_f & -\frac{s}{c} Q_f & 0 \\ \hline
\earr
\eeq
}

{\item
{\samepage
$\Shat \bar{F}F$ coupling:
\beq \label{fr:sff}
\barr{l}
\framebox{
\begin{picture}(96,82)(0,-4)
\put(75,65){\makebox(10,20)[bl]{$\bar{F}_{1}$}}
\put(0,42){\makebox(10,20)[bl]{$\Shat$}}
\put(75,-2){\makebox(10,20)[bl]{$F_{2}$}}
\put(48,36){\circle*{4}}
\put(-2,36){\usebox{\Sr}}
\put(48,12){\usebox{\Ftbr}}
\end{picture} }
\earr
\renewcommand{\arraystretch}{2.0}
\barr{l}
\disp = ie\Bigl[ C_{\rL}\omega_{-}
  \Bigl(1 + \frac{\delta m_{F_1}}{m_{F_1}}
  + \frac{1}{2} \delta Z^{\rL}_{F_1}
  + \frac{1}{2} \delta Z^{\rR}_{F_1}\Bigr) \\
\disp \mbox{} \quad + C_{\rR}\omega_{+}
  \Bigl(1 + \frac{\delta m_{F_2}}{m_{F_2}}
  + \frac{1}{2} \delta Z^{\rL}_{F_1} +
  \frac{1}{2} \delta Z^{\rR}_{F_2}\Bigr)
\Bigr]
\earr
\eeq
}
\renewcommand{\arraystretch}{1.4}
with the actual values of $\Shat$, $\bar{F}_{1}$, $F_{2}$ and
$C_{\rR}$, $C_{\rL}$
\beq \label{fr:Csff}
\barr{|r||c|c|c|c|} \hline
\Shat\bar{F}_{1} F_{2} & \Hhat\bar{f} f & \chihat\bar{f} f &
\phihat^+\bar{f}_u f_d & \phihat^-\bar{f}_d f_u \\ \hline
C_{\rL} & -\frac{1}{2 s} \frac{m_f}{\MW} &
-i \frac{1}{2 s} 2 I_{\rw,f}^3 \frac{m_{f}}{\MW} &
+\frac{1}{\sqrt{2} s} \frac{m_{f_u}}{\MW} &
-\frac{1}{\sqrt{2} s} \frac{m_{f_d}}{\MW} \\ \hline
C_{\rR} & -\frac{1}{2 s} \frac{m_f}{\MW} &
i \frac{1}{2 s} 2 I_{\rw,f}^3 \frac{m_f}{\MW} &
-\frac{1}{\sqrt{2} s} \frac{m_{f_d}}{\MW} &
+\frac{1}{\sqrt{2} s} \frac{m_{f_u}}{\MW} \\ \hline
\earr
\eeq
}%

\end{itemize}

Note that in contrast to the conventional formalism no counterterms are
needed for the $\Zhat\Ahat\Hhat\Hhat$, $\Zhat\Ahat\chihat\chihat$,
$\Ahat\chihat\Hhat$ and $\Hhat\Zhat\Ahat$ couplings.

We now consider the Feynman rules for vertices
containing quantum fields. We treat these vertices in lowest
order, i.e.~we do not list the counterterms explicitly.
As mentioned above, all lowest-order vertices involving
fermions have the usual form.
Since the gauge-fixing term is quadratic in the quantum fields,
apart from vertices involving ghost fields only vertices
containing exactly two quantum fields differ from the
conventional ones. Thus, the other vertices involving quantum fields
have in lowest order the same form as the pure background-field
vertices given above.
Their insertions can be obtained from the ones listed for the
pure background-field vertices by forming all possible
combinations of quantum and background fields,
e.g.~one infers $\What^+W^-AZ, W^+\What^-AZ, W^+W^-\Ahat Z$ and
$W^+W^-A\Zhat$ as the possible insertions for the $\Vhat VVV$
coupling corresponding to $\What^+\What^-\Ahat\Zhat$.

Some of the vertices containing two quantum fields also have the
usual Feynman rules. These are $\Vhat\Vhat SS$, $\Shat\Shat V V$,
$\Vhat SS$ and $\Shat VV$. In the following we list those couplings
for which the generic form or actual insertion differs from
the ones in the conventional formalism. Note that some of the
insertions appearing in the conventional couplings have no counterpart here.
We list only the non-vanishing insertions.

\begin{itemize}
{\item
{\samepage
$\Vhat\Vhat VV$ coupling:\\
The $\Vhat\Vhat VV$ coupling has two generic forms
depending on the actual insertions, %We find
\beq
\Vhat_{1, \mu}\Vhat_{2, \nu} V_{3, \rho}V_{4, \si} : i e^2 C \Bigl[
2 g_{\mu\nu} g_{\rho\si} - g_{\mu\si} g_{\nu\rho} (1 - \frac{1}{\xiQ})
- g_{\mu\rho} g_{\nu\si} (1 - \frac{1}{\xiQ}) \Bigr]
\eeq
}
for the insertions
\beq
\barr{|r||c|c|c|c|} \hline
\Vhat_1\Vhat_2 V_3V_4 & \What^\pm\What^\pm W^\mp W^\mp &
\Zhat\Zhat W^+W^- & \Ahat\Zhat W^+W^- & \Ahat\Ahat W^+W^-\\
\mbox{}       & & \What^+\What^-ZZ &
\What^+\What^-AZ & \What^+\What^-AA \\ \hline
C & \frac{1}{s^2} & -\frac{c^2}{s^2} & \frac{c}{s} & -1\\ \hline
\earr
\eeq
and
\beq
\Vhat_{1, \mu}\Vhat_{2, \nu} V_{3, \rho}V_{4, \si} : i e^2 C \Bigl[
2 g_{\mu\rho} g_{\nu\si} - g_{\mu\nu} g_{\rho\si} -
g_{\mu\si} g_{\nu\rho} (1 + \frac{1}{\xiQ}) \Bigr]
\eeq
for the insertions
\beq
\barr{|r||c|c|c|c|} \hline
\Vhat_1\Vhat_2 V_3V_4 & \What^+\What^-W^+W^- &
\What^{\pm}\Zhat W^{\mp}Z & % \What^{\pm}\Zhat W^{\mp}A,
\What^{\pm}\Ahat W^{\mp}Z & \What^{\pm}\Ahat W^{\mp}A\\ %\hline
 & & & \What^{\pm}\Zhat W^{\mp}A & \\ \hline
C & \frac{1}{s^2} & -\frac{c^2}{s^2} & \frac{c}{s} & -1\\ \hline
\earr
\eeq
}

{\item
{\samepage
$\Vhat VV$ coupling: %\\
\beqar
\lefteqn{\Vhat_{1, \mu}(k_1)V_{2, \nu}(k_2)V_{3, \rho}(k_3) :}\no
&&  - i e C \Bigl[
g_{\nu\rho} (k_3 - k_2)_{\mu} +
g_{\mu\nu} (k_2 - k_1 + \frac{k_3}{\xiQ})_{\rho} +
g_{\rho\mu} (k_1 - k_3 - \frac{k_2}{\xiQ})_{\nu} \Bigr]
\eeqar
}
with the actual values of $\Vhat_1$, $V_2$, $V_3$ and $C$
\beq
\barr{|r||c|c|} \hline
\Vhat_1 V_2 V_3 & \Ahat W^+W^-, \What^+ W^- A, \What^- A W^+ &
\Zhat W^+W^-, \What^+ W^- Z, \What^- Z W^+\\ \hline
C & 1 & - \frac{c}{s}\\ \hline
\earr
\eeq
}

{\item
{\samepage
\noindent
$\Shat\Shat SS$ coupling: %\\
\beq
\Shat_1 \Shat_2 S_3 S_4 : i e^2 C
\eeq
}
with the actual values of $\Shat_1$, $\Shat_2$, $S_3$, $S_4$ and $C$
\bma
\barr{|r||c|c|c|c|} \hline
\Shat_1 \Shat_2 S_3 S_4 & \Hhat\Hhat HH & \Hhat\Hhat\chi\chi &
\Hhat\chihat H\chi & \phihat^+\phihat^-HH, \Hhat\Hhat\phi^+\phi^- \\
&  \chihat\chihat\chi\chi & \chihat\chihat HH &
& \phihat^+\phihat^-\chi\chi, \chihat\chihat\phi^+\phi^- \\ \hline
C & -\frac{3}{4 s^2} \frac{\MH^2}{\MW^2} &
- \frac{1}{4 s^2} \frac{\MH^2}{\MW^2} - \frac{\xiQ}{2 c^2 s^2} &
- \frac{1}{4 s^2} \frac{\MH^2}{\MW^2} + \frac{\xiQ}{4 c^2 s^2} &
- \frac{1}{4 s^2} \frac{\MH^2}{\MW^2} - \frac{\xiQ}{2 s^2} \\ \hline
\earr
\ema
and
\beq
\barr{|r||c|c|c|c|} \hline
\Shat_1 \Shat_2 S_3 S_4 & \phihat^{\pm}\Hhat\phi^{\mp}H
& \phihat^+\phihat^-\phi^+\phi^- &
\phihat^{\pm}\phihat^{\pm}\phi^{\mp}\phi^{\mp} &
\phihat^{\pm}\Hhat\phi^{\mp}\chi\\
& \phihat^{\pm}\chihat\phi^{\mp}\chi
& & & \phihat^{\mp}\chihat\phi^{\pm}H  \\ \hline
C & - \frac{1}{4 s^2} \frac{\MH^2}{\MW^2} + \frac{\xiQ}{4 s^2} &
- \frac{1}{2 s^2} \frac{\MH^2}{\MW^2} - \frac{\xiQ}{4 c^2 s^2} &
- \frac{1}{2 s^2} \frac{\MH^2}{\MW^2} + \frac{\xiQ}{2 c^2 s^2}
& \mp \frac{i \xiQ}{4 c^2} \\ \hline
\earr
\eeq
}

{\item
{\samepage
$\Shat SS$ coupling:
\beq
\Shat_1 S_2 S_3 : i e C
\eeq
}
with the actual values of $\Shat_1$, $S_2$, $S_3$ and $C$
\bma
\barr{|r||c|c|c|c|c|c|} \hline
\Shat_1 S_2 S_3 & \Hhat HH & \Hhat\chi\chi &
\chihat H \chi & \Hhat \phi^+\phi^- \\ \hline
C & -\frac{3}{2 s} \frac{\MH^2}{\MW} &
-\frac{1}{2 s} \frac{\MH^2}{\MW} - \xiQ \frac{\MW}{c^2 s} &
-\frac{1}{2 s} \frac{\MH^2}{\MW} + \xiQ \frac{\MW}{2 c^2 s} &
-\frac{1}{2 s} \frac{\MH^2}{\MW} - \xiQ \frac{\MW}{s} \\ \hline
\earr
\ema
and
\beq
\barr{|r||c|c|} \hline
\Shat_1 S_2 S_3 & \phihat^{\pm}\phi^{\mp}H
& \phihat^{\pm}\phi^{\mp}\chi\\ \hline
C & -\frac{1}{2 s} \frac{\MH^2}{\MW} + \xiQ \frac{\MW}{2 s}
& \mp i \xiQ \MW\frac{s}{2 c^2}\\
\hline
\earr
\eeq
}

{\item
{\samepage
$\Vhat V \Shat S$ coupling:
\beq
\Vhat_{1, \mu} V_{2, \nu} \Shat_1 S_2 : i e^2 g_{\mu\nu} C
\eeq
}
with the actual values of $\Vhat_1$, $V_2$, $\Shat_1$, $S_2$ and $C$
\bma
\barr{|r||c|c|c|c|c|c|c|} \hline
\Vhat_1 V_2\Shat_1 S_2  & \Zhat Z\Hhat H
& \What^{\pm}W^{\mp}\Hhat H & \What^{\pm}W^{\mp}\phihat^{\mp}\phi^{\pm}
& \Ahat A \phihat^{\pm}\phi^{\mp} & \Zhat A \phihat^{\pm}\phi^{\mp}
& \Zhat Z \phihat^{\pm}\phi^{\mp} \\
& \Zhat Z\chihat\chi & \What^{\pm}W^{\mp}\chihat\chi & & &
\Ahat Z\phihat^{\pm}\phi^{\mp} &
\\ \hline
C & \frac{1}{2 c^2 s^2} & \frac{1}{2 s^2} & \frac{1}{s^2} & 2
& -\frac{c^2 - s^2}{c s} & \frac{(c^2 - s^2)^2}{2 c^2 s^2} \\ \hline
\earr
\ema
and
\bma
\barr{|c|c|c|c|c|c|c|} \hline
\Vhat_1 V_2\Shat_1 S_2 &
\What^{\pm}A\Hhat\phi^{\mp} &
\What^{\pm}A\chihat\phi^{\mp} &
\What^{\pm}Z\phihat^{\mp}H &
\What^{\pm}Z\phihat^{\mp}\chi &
\What^{\pm}Z\Hhat\phi^{\mp} &
\What^{\pm}Z\chihat\phi^{\mp} \\
& \Ahat W^{\pm}\phihat^{\mp}H &
\Ahat W^{\pm}\phihat^{\mp}\chi &
\Zhat W^{\pm}\Hhat \phi^{\mp} &
\Zhat W^{\pm}\chihat \phi^{\mp} &
\Zhat W^{\pm}\phihat^{\mp}H &
\Zhat W^{\pm}\phihat^{\mp}\chi
\\ \hline
C & -\frac{1}{s} & \mp\frac{i}{s} & -\frac{1}{2 c s^2} & \mp\frac{i}{2 c s^2} &
\frac{c^2 - s^2}{2 c s^2} & \pm i \frac{c^2 - s^2}{2 c s^2}
\\ \hline
\earr
\ema
and
\beq
\barr{|c|c|c|c|c|c|c|} \hline
\Vhat_1 V_2\Shat_1 S_2 &
\What^\pm W^\mp\chihat H \\
&\What^\mp W^\pm\Hhat\chi \\ \hline
C & \pm\frac{i}{2 s^2} \\ \hline
\earr
\eeq
}

{\item
{\samepage
$V\Shat S$ coupling:
\beq
V_{\mu} \Shat_1(k_1) S_2(k_2) : i e C 2 k_{1, \mu}
\eeq
}
with the actual values of $V$, $\Shat_1$, $S_2$ and $C$
\beqar
\barr{|r||c|c|c|c|c|c|c|} \hline
V\Shat_1 S_2 & Z\chihat H & Z\Hhat\chi & A\phihat^\pm\phi^\mp &
Z\phihat^\pm\phi^\mp & W^{\pm}\phihat^{\mp} H, W^{\mp}\Hhat\phi^\pm &
W^{\pm}\phihat^{\mp}\chi & W^{\pm}\chihat\phi^{\mp} \\ \hline
C & -\frac{i}{2 c s}  & \frac{i}{2 c s} & \mp1 & \pm\frac{c^2 - s^2}{2 c s} &
\mp \frac{1}{2 s} & -\frac{i}{2 s} & \frac{i}{2 s} \\ \hline
\earr\no
\eeqar
}

{\item
{\samepage
$S \Vhat V$ coupling:
\beq
S \Vhat_{1, \mu} V_{2, \nu} : i e g_{\mu \nu} C
\eeq
}
with the actual values of $S$, $\Vhat_1$, $V_2$ and $C$
\beq
\barr{|r||c|c|c|c|c|c|c|} \hline
S \Vhat_1 V_2 & H\Zhat Z & H\What^{\pm}W^{\mp} & \chi\What^\pm W^\mp &
\phi^{\pm}\What^{\mp}A & \phi^{\pm}\What^{\mp}Z & \phi^{\pm}\Zhat W^{\mp}
\\ \hline
C & \frac{1}{c^2 s} \MW & \frac{1}{s} \MW & \mp\frac{i}{s} \MW &
- 2 \MW & \frac{c^2 - s^2}{c s} \MW & - \frac{1}{c s} \MW
\\ \hline
\earr
\eeq
}
\end{itemize}

Next, we list the Feynman rules for couplings involving ghost
fields. As above, pure quantum-field vertices have the usual
Feynman rules. % (see e.g.~\citere{Dehab}).
\nobreak
\begin{itemize}

{\item
{\samepage
$\Vhat\bar GG$ coupling:
\beq \label{fr:vgg}
\barr{l}
\framebox {
\begin{picture}(96,82)(0,-4)
\put(67,65){\makebox(10,20)[bl]{$\bar{G_{1}},k_{1}$}}
\put(0,42){\makebox(10,20)[bl]{$\Vhat_{\mu}$}}
\put(67,-4){\makebox(10,20)[bl]{$G_{2},k_2$}}
\put(48,36){\circle*{4}}
\put(0,34){\usebox{\Vr}}
\put(48,12){\usebox{\Gtbr}}
\end{picture} }
\earr
\barr{l}
= ie(k_{1}-k_2)_{\mu}C
\earr
\eeq
}
{\samepage
with the actual values of $\Vhat$, $\bar{G}_{1}$, $G_{2}$ and $C$
\beq
\label{fr:Cvgg}
\barr{|r||c|c|} \hline
\Vhat\bar{G_{1}}G_2 & \Ahat\bar{u}^{\pm} u^{\pm},
\What^{\pm} \bar{u}^{A} u^{\mp}, \What^{\mp} \bar{u}^{\mp} u^{A} &
\Zhat\bar{u}^{\pm} u^{\pm}, \What^{\pm} \bar{u}^{Z} u^{\mp},
\What^{\mp} \bar{u}^{\mp} u^{Z}  \\ \hline
C & \pm{1} & \mp\frac{c}{s} \\ \hline
\earr
\eeq
}
}

{\item
{\samepage
$V\bar GG$ coupling:
\beq
V_{\mu} \bar{G}_1 G_2 : ie k_{1,\mu} C
\eeq
}
with the actual values of $V$, $\bar{G}_{1}$, $G_{2}$ and $C$
as given in \refeq{fr:Cvgg}.
}

{\item
{\samepage
$\Vhat\Vhat\bar GG$ coupling:
\beq \label{fr:vvgg}
\barr{l}
\framebox {
\begin{picture}(96,82)(0,-4)
\put(75,65){\makebox(10,20)[bl]{$\bar{G_{1}}$}}
\put(75,-4){\makebox(10,20)[bl]{$G_{2}$}}
\put(6,65){\makebox(10,20)[bl]{$\Vhat_{1,\mu}$}}
\put(6,-4){\makebox(10,20)[bl]{$\Vhat_{2,\nu}$}}
\put(48,36){\circle*{4}}
\put(16,12){\usebox{\Vtr}}
\put(16,36){\usebox{\Vbr}}
\put(48,12){\usebox{\Gtbr}}
\end{picture} }
\earr
\barr{l}
= ie^2g_{\mu\nu}C
\earr
\eeq
}
with the actual values of $\Vhat_1$, $\Vhat_2$, $\bar{G}_{1}$, $G_{2}$ and $C$
\begin{subequations}
\beqar
\barr{|r||c|c|c|c|} \hline
\Vhat_1\Vhat_2\bar{G_{1}}G_2 &
\What^\pm\What^\pm\bar{u}^{\pm}u^{\mp} &
\What^+\What^-\bar{u}^{A}u^{A}& \What^+\What^-\bar{u}^{A}u^{Z},
\Ahat\Zhat\bar{u}^{\pm} u^{\pm} & \What^+\What^-\bar{u}^{Z}u^{Z}
\\
&&\Ahat\Ahat\bar{u}^{\pm} u^{\pm} & \What^+\What^-\bar{u}^{Z}u^{A} \hfill
& \Zhat\Zhat\bar{u}^{\pm} u^{\pm} \\\hline
C & -\frac{2}{s^2} & 2 & -2\frac{c}{s} & 2\frac{c^2}{s^2} \\ \hline
\earr \no
\eeqar
and
\beqar
\barr{|r||c|c|c|c|} \hline
\Vhat_1\Vhat_2\bar{G_{1}}G_2 &
\What^+\What^-\bar{u}^{\pm}u^{\pm} &
\Ahat\What^\pm\bar{u}^{\pm} u^{A} & \Zhat\What^\pm\bar{u}^{\pm} u^{A} ,
\Ahat\What^\pm\bar{u}^{\pm} u^{Z} & \Zhat\What^\pm\bar{u}^{\pm} u^{Z}
\\
&&\Ahat\What^\pm\bar{u}^{A} u^{\mp} & \Zhat\What^\pm\bar{u}^{A} u^{\mp} ,
\Ahat\What^\pm\bar{u}^{Z} u^{\mp} & \Zhat\What^\pm\bar{u}^{Z} u^{\mp}
\\\hline
C &  \frac{1}{s^2} & -1 & \frac{c}{s} & -\frac{c^2}{s^2} \\ \hline
\earr\no  \label{fr:Cvvgg}
\eeqar
\end{subequations}
}

{\item
{\samepage
$\Vhat V\bar GG$ coupling:
\beq
\Vhat_{1, \mu} V_{2, \nu} \bar{G}_1 G_2 : i e^2 g_{\mu \nu} C
\eeq
}
with the actual values of $\Vhat_1$, $\Vhat_2$, $\bar{G}_{1}$, $G_{2}$ and $C$
\bma
\barr{|r||c|c|c|c|} \hline
\Vhat_1 V_2\bar{G_{1}}G_2 &
\What^\pm W^\pm\bar{u}^{\pm}u^{\mp} &
\What^\pm W^\mp\bar{u}^{A}u^{A}& \What^\pm W^\mp\bar{u}^{A}u^{Z},
\Ahat Z\bar{u}^{\pm} u^{\pm} & \What^\pm W^\mp\bar{u}^{Z}u^{Z}
\\
&&\Ahat A\bar{u}^{\pm} u^{\pm} & \What^\pm\What^\mp\bar{u}^{Z}u^{A},
\Zhat A\bar{u}^{\pm} u^{\pm} & \Zhat Z\bar{u}^{\pm} u^{\pm} \\\hline
C  & -\frac{1}{s^2} & 1 & -\frac{c}{s} & \frac{c^2}{s^2} \\ \hline
\earr
\ema
and
\beq
\barr{|r||c|c|c|c|} \hline
\Vhat_1 V_2\bar{G_{1}}G_2 &
\What^\pm W^\mp\bar{u}^{\pm}u^{\pm} &
\Ahat W^\pm\bar{u}^{\pm} u^{A} & \Zhat W^\pm\bar{u}^{\pm} u^{A} ,
\Ahat W^\pm\bar{u}^{\pm} u^{Z} & \Zhat W^\pm\bar{u}^{\pm} u^{Z}
\\
&&\What^\pm A\bar{u}^{A} u^{\mp} & \What^\pm Z\bar{u}^{A} u^{\mp} ,
\What^\pm A\bar{u}^{Z} u^{\mp} & \What^\pm Z\bar{u}^{Z} u^{\mp}
\\\hline
C  &  \frac{1}{s^2} & -1 & \frac{c}{s} & -\frac{c^2}{s^2} \\ \hline
\earr
\eeq
}

{\item
{\samepage
$\Shat\bar GG$ coupling:
\beq \label{fr:sgg}
\barr{l}
\framebox {
\begin{picture}(96,82)(0,-4)
\put(75,65){\makebox(10,20)[bl]{$\bar{G_{1}}$}}
\put(0,42){\makebox(10,20)[bl]{$\Shat$}}
\put(75,-4){\makebox(10,20)[bl]{$G_{2}$}}
\put(48,36){\circle*{4}}
\put(0,36){\usebox{\Sr}}
\put(48,12){\usebox{\Gtbr}}
\end{picture} }
\earr
\barr{l}
= ieC\xi_Q
\earr
\eeq
}
with the actual values of $\Shat$, $\bar{G}_{1}$, $G_{2}$ and $C$
\beq
\label{fr:Csgg}
\barr{|r||c|c|c|c|c|c|} \hline
\Shat\bar{G_{1}}G_2 & \Hhat\bar{u}^{Z}u^{Z} & \Hhat\bar{u}^{\pm} u^{\pm}
& \phihat^{\pm} \bar{u}^{\pm} u^{A}, \phihat^{\pm} \bar{u}^{A} u^{\mp} &
  \phihat^{\pm} \bar{u}^{\pm} u^{Z}, \phihat^{\pm} \bar{u}^{Z} u^{\mp}
 \\ \hline
C & -\frac{1}{c^{2}s}\MW & -\frac{1}{s}\MW &
\MW & \frac{s}{c}\MW \\ \hline
\earr
\eeq
}

{\item
{\samepage
$S\bar GG$ coupling:
\beq
S \bar{G}_1 G_2 : ieC\xi_Q
\eeq
}
with the actual values of $S$, $\bar{G}_{1}$, $G_{2}$ and $C$
\beq
\barr{|r||c|c|c|c|c|c|} \hline
S\bar{G_{1}}G_2 & H\bar{u}^{Z}u^{Z} & H\bar{u}^{\pm} u^{\pm}
& \chi\bar{u}^{\pm} u^{\pm}         & \phi^{\pm} \bar{u}^{\pm} u^{A}
& \phi^{\pm} \bar{u}^{\pm} u^{Z} & \phi^{\pm} \bar{u}^{Z} u^{\mp}
 \\ \hline
C & -\frac{1}{2c^{2}s}\MW & -\frac{1}{2s}\MW &  \mp\frac{i}{2s}\MW &
\MW & -\frac{c^2-s^2}{2cs}\MW  & \frac{1}{2cs}\MW \\ \hline
\earr
\eeq
}

{\item
{\samepage
$\Shat\Shat\bar GG$ coupling:
\beq \label{fr:ssgg}
\barr{l}
\framebox {
\begin{picture}(96,82)(0,-4)
\put(75,65){\makebox(10,20)[bl]{$\bar{G_{1}}$}}
\put(9,65){\makebox(10,20)[bl]{$\Shat_{1}$}}
\put(9,-4){\makebox(10,20)[bl]{$\Shat_{2}$}}
\put(75,-4){\makebox(10,20)[bl]{$G_{2}$}}
\put(48,36){\circle*{4}}
\put(16,12){\usebox{\Str}}
\put(16,36){\usebox{\Sbr}}
\put(48,12){\usebox{\Gtbr}}
\end{picture} }
\earr
\barr{l}
= ie^2C\xi_Q
\earr
\eeq
}
with the actual values of $\Shat_1$, $\Shat_2$, $\bar{G}_{1}$, $G_{2}$ and $C$
\bma
\barr{|r||c|c|c|c|c|c|} \hline
\Shat_1 \Shat_2\bar{G_{1}}G_2
& \Hhat\Hhat\bar{u}^{Z}u^{Z} & \Hhat\Hhat\bar{u}^{\pm} u^{\pm}\!
, \phihat^+\phihat^- \bar{u}^{\pm} u^{\pm}
& \phihat^+\phihat^-\bar{u}^{A}u^{A}
& \phihat^+\phihat^-\bar{u}^{A}u^{Z}
& \phihat^+\phihat^-\bar{u}^{Z}u^{Z}  \\
& \chihat\chihat \bar{u}^{Z}u^{Z} & \chihat\chihat \bar{u}^{\pm} u^{\pm}
\hfill  & & \phihat^+\phihat^-\bar{u}^{Z}u^{A}  &
\\ \hline
C & -\frac{1}{2c^{2}s^2} & -\frac{1}{2s^2} &
-2 & \frac{c^2-s^2}{cs} & - \frac{(c^2-s^2)^2}{2c^2s^2}  \\ \hline
\earr
\ema
and
\beq
\label{fr:Cssgg}
\barr{|r||c|c|c|c|c|} \hline
\Shat_1\Shat_2\bar{G_{1}}G_2
& \Hhat \phihat^\pm \bar{u}^{\pm}u^{A}
& \chihat \phihat^\pm \bar{u}^{\pm} u^{A}
& \Hhat \phihat^\pm \bar{u}^{\pm}u^{Z}
& \chihat \phihat^\pm \bar{u}^{\pm} u^{Z} \\
& \phihat^\pm \Hhat \bar{u}^{A}u^{\mp}
& \phihat^\pm \chihat \bar{u}^{A} u^{\mp}
& \phihat^\pm \Hhat \bar{u}^{Z}u^{\mp}
& \phihat^\pm \chi \bar{u}^{Z} u^{\mp}  \\ \hline
C & \frac{1}{2s} & \mp\frac{i}{2s}
& \frac{1}{2c} & \mp i\frac{1}{2c}  \\ \hline
\earr
\eeq
}

{\item
{\samepage
$\Shat S \bar GG$ coupling:
\beq
\Shat_{1} S_{2} \bar{G}_1 G_2 : i e^2  C \xi_Q
\eeq
}
with the actual values of $\Shat_1$, $S_2$, $\bar{G}_{1}$, $G_{2}$ and $C$
\bma
\barr{|r||c|c|c|c|c|c|c|} \hline
\Shat_1 S_2\bar{G_{1}}G_2
& \Hhat H \bar{u}^{Z}u^{Z} & \Hhat H \bar{u}^{\pm} u^{\pm}
& \phihat^{\pm}\phi^\mp \bar{u}^{\pm} u^{\pm}
& \phihat^{\pm}\phi^\mp\bar{u}^{A}u^{A}
& \phihat^{\pm}\phi^\mp\bar{u}^{A}u^{Z}
& \phihat^{\pm}\phi^\mp\bar{u}^{Z}u^{Z} \\
& \chihat\chi \bar{u}^{Z}u^{Z} & \chihat\chi \bar{u}^{\pm} u^{\pm}
& & & \phihat^{\pm}\phi^\mp\bar{u}^{Z}u^{A} & \\ \hline
C & -\frac{1}{4c^{2}s^2} & -\frac{1}{4s^2} & -\frac{1}{2s^2} & -1
& \frac{c^2-s^2}{2cs} & -\frac{(c^2-s^2)^2}{4c^2s^2} \\ \hline
\earr
\ema
and
\bma
\barr{|r||c|c|c|c|c|c|} \hline
\Shat_1 S_2\bar{G_{1}}G_2
& \Hhat \phi^\pm \bar{u}^{\pm}u^{A}
& \chihat \phi^\pm \bar{u}^{\pm} u^{A}
& \Hhat \phi^\pm \bar{u}^{\pm}u^{Z}
& \Hhat \phi^\pm \bar{u}^{Z}u^{\mp}
& \chihat \phi^\pm \bar{u}^{\pm} u^{Z}
& \chihat \phi^\pm \bar{u}^{Z} u^{\mp} \\
& \phihat^\pm H \bar{u}^{A}u^{\mp}
& \phihat^\pm \chi \bar{u}^{A} u^{\mp}
& \phihat^\pm H \bar{u}^{Z}u^{\mp}
& \phihat^\pm H \bar{u}^{\pm}u^{Z}
& \phihat^\pm \chi \bar{u}^{Z} u^{\mp}
& \phihat^\pm \chi \bar{u}^{\pm} u^{Z} \\ \hline
C & \frac{1}{2s} & \mp\frac{i}{2s}
& -\frac{c^2-s^2}{4cs^{2}} & \frac{1}{4cs^2}
& \pm i\frac{c^2-s^2}{4cs^2} & \mp\frac{i}{4cs^2}  \\ \hline
\earr
\ema
and
\beq
\barr{|r||c|c|c|c|c|c|} \hline
\Shat_1 S_2\bar{G_{1}}G_2
& \Hhat\chi\bar{u}^{\pm} u^{\pm} \\
& \chihat H\bar{u}^{\mp} u^{\mp} \\ \hline
C & \mp\frac{i}{4s^2} \\ \hline
\earr
\eeq
}
\end{itemize}

Finally, we give the quantum-field propagators:
\begin{itemize}
\item
gauge bosons $V = A$, $Z$, $W$  ($\MA = 0$)
\beq
\barr{l}
\framebox{
\begin{picture}(90,24)
\put(33,16){\makebox(20,10)[bl] {$k$}}
\put(-1,9){\makebox(12,10)[bl] {$V_{\mu}$}}
\put(67,9){\makebox(12,10)[bl] {$V_{\nu}$}}
\put(15,10){\usebox{\Vr}}
\put(15,12){\circle*{4}}
\put(63,12){\circle*{4}}
\end{picture} }
\earr
\barr{l}
\disp = -i\left[\frac{g_{\mu\nu}}{k^{2}-M_V^{2}}
     -\frac{(1-\xi_Q)k_\mu k_\nu}{(k^{2}-M_V^{2})(k^{2}-\xi_Q M_V^{2})}\right]
\earr
\eeq

\item
Faddeev--Popov ghosts $G = u^{A}$, $u^{Z}$, $u^{\pm}$
($M_{u^{A}}=0$, $M_{u^{Z}} = \xi_Q\MZ$, $M_{u^{\pm}} = \xi_Q\MW$)
\beq
\barr{l}
\framebox{
\begin{picture}(90,24)
\put(34,16){\makebox(20,10)[bl] {$k$}}
\put(0,9){\makebox(12,10)[bl] {$G$}}
\put(67,9){\makebox(12,10)[bl] {$\bar{G}$}}
\put(15,12){\usebox{\Gr}}
\put(15,12){\circle*{4}}
\put(63,12){\circle*{4}}
\end{picture} }
\earr
\barr{l}
\disp = \frac{i}{k^{2}-M_{G}^{2}}
\earr
\eeq

\item
scalar fields $S = H$, $\chi$, $\phi$
($M_\chi = \xi_Q \MZ$, $M_\phi = \xi_Q \MW$)
\beq
\barr{l}
\framebox{
\begin{picture}(90,24)
\put(34,16){\makebox(20,10)[bl] {$k$}}
\put(0,9){\makebox(12,10)[bl] {$S$}}
\put(67,9){\makebox(12,10)[bl] {$S$}}
\put(15,12){\usebox{\Sr}}
\put(15,12){\circle*{4}}
\put(63,12){\circle*{4}}
\end{picture} }
\earr
\barr{l}
\disp= \frac{i}{k^{2}-M_{S}^{2}}
\earr
\eeq

\item
fermion fields $F = f$
\beq
\barr{l}
\framebox{
\begin{picture}(90,24)
\put(34,17){\makebox(20,10)[bl] {$p$}}
\put(0,9){\makebox(12,10)[bl] {$F$}}
\put(67,9){\makebox(12,10)[bl] {$\bar{F}$}}
\put(15,12){\usebox{\Fr}}
\put(15,12){\circle*{4}}
\put(63,12){\circle*{4}}
\end{picture} }
\earr
\barr{l}
\disp= \frac{i(\ps + m_{F})}{p^{2}-m_{F}^{2}}
\earr
\eeq

\end{itemize}

\end{document}